\documentclass[bibyear]{aa}
\usepackage{graphicx}
\usepackage{natbib}
\usepackage{booktabs}
\bibpunct{(}{)}{;}{a}{}{,}
\usepackage{txfonts}
\usepackage{soul}
\usepackage{subcaption}
\usepackage{float}
\usepackage[colorlinks=true,linkcolor=blue,citecolor=blue,filecolor=blue]{hyperref}
\usepackage[normalem]{ulem}

\newcommand{\linkorcid}[1]{\href{https://orcid.org/#1}{\includegraphics[width=8pt]{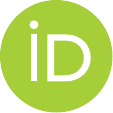}}}

\begin{document}

   \title{Back from the dead: AT2019aalc as a candidate repeating TDE in an AGN}
    \titlerunning{AT2019aalc as a repeating TDE in an AGN}

   \author{Patrik Mil\'an Veres \inst{\ref{rub}}\linkorcid{0000-0002-9553-2987}
          \and
          Anna Franckowiak\inst{\ref{rub},\ref{rapp}}\linkorcid{0000-0002-5605-2219}
          \and
          Sjoert van Velzen \inst{\ref{leiden}}\linkorcid{0000-0002-3859-8074}
          \and
          Bjoern Adebahr\inst{\ref{rub}}
          \and
          Sam Taziaux\inst{\ref{rub}}\linkorcid{0009-0001-6908-2433}
          \and
          Jannis Necker\inst{\ref{desy}, \ref{hu}}\linkorcid{0000-0003-0280-7484}
          \and
          Robert Stein\inst{\ref{caltech}}\linkorcid{0000-0003-2434-0387}
          \and
          Alexander Kier\inst{\ref{rub}}\linkorcid{0009-0009-7572-3645}
          \and 
          Ancla M\"{u}ller \inst{\ref{rub}}
          \and
          Dominik J.\ Bomans\inst{\ref{rub},\ref{rapp}}\linkorcid{0000-0001-5126-5365}
          \and
          Nuria Jordana-Mitjans \inst{\ref{rub}}
          \and
          Marek Kowalski
          \inst{\ref{desy}}\linkorcid{0000-0001-8594-8666}
          \and
          Erica Hammerstein
          \inst{\ref{umd}, \ref{gsfc}, \ref{cresst}}
          \and
          Elena Marci-Boehncke \inst{\ref{rub}}\linkorcid{0009-0005-2161-1403}
          \and
          Simeon Reusch \inst{\ref{desy},\ref{hu}}\linkorcid{0000-0002-7788-628X}
          \and
          Simone Garrappa\inst{\ref{wis}}\linkorcid{0000-0003-2403-4582}
          \and
          Sam Rose\inst{\ref{caltech}}\linkorcid{0000-0003-4725-4481}
          \and
          Kaustav Kashyap Das \inst{\ref{cahill}}\linkorcid{0000-0001-8372-997X}
          }

   \institute{Ruhr University Bochum, Faculty of Physics and Astronomy, Astronomical Institute (AIRUB), Universitätsstraße 150, 44801 Bochum, Germany
              \email{veres@astro.ruhr-uni-bochum.de} \label{rub}
        \and 
             Ruhr Astroparticle and Plasma Physics Center (RAPP Center) \label{rapp}
        \and 
             Leiden Observatory, Leiden University, Postbus 9513, 2300 RA Leiden, The Netherlands \label{leiden}
         \and
             Deutsches Elektronen-Synchrotron (DESY), Platanenallee 6, D-15378 Zeuthen, Germany \label{desy}
        \and
            Institut fur Physik, Humboldt-Universität zu Berlin, D-12489 Berlin, Germany \label{hu}
        \and
            Division of Physics, Mathematics, and Astronomy, California Institute of Technology, Pasadena, CA 91125, USA
            \label{caltech}
        \and
            University of Maryland, Department of Astronomy, 4296 Stadium Dr, College Park, MD 20742, USA \label{umd}
        \and
            NASA Goddard Space Flight Center, Astrophysics Science Division, 8800 Greenbelt Rd, Greenbelt, MD 20771, USA \label{gsfc}
        \and
            Center for Research and Exploration in Space Science and Technology, NASA/GSFC, Greenbelt, MD 20771, USA \label{cresst}
        \and 
            Department of Particle Physics and Astrophysics, Weizmann Institute of Science, 76100 Rehovot, Israel.
            \label{wis}
        \and
           Cahill Center for Astrophysics, California Institute of Technology, MC 249-17, 1200 E California Boulevard, Pasadena, CA, 91125, USA \label{cahill}
            }

   \date{}

  \abstract 
   {To date, three nuclear transients have been associated with high-energy neutrino events. These transients are generally thought to be powered by tidal disruptions of stars (TDEs) by massive black holes. However, AT2019aalc, hosted in a Seyfert-1 galaxy, was not yet classified due to a lack of multiwavelength observations. Interestingly, the source has re-brightened 4 years after its discovery.}
   {Our aim is to constrain the physical mechanism responsible for the second optical flare, which may also provide clues to the origin of the initial event.}
   {We conducted a multi-wavelength monitoring program (from radio to X-rays) of AT2019aalc during its re-brightening in 2023/2024.}
   {The observations revealed multiple X-ray flares during the second optical flaring episode of the transient and a uniquely bright UV counterpart. The second flare, similarly to the first one, is also accompanied by IR dust echo emission. A long-term radio flare is found with an inverted spectrum. Optical spectroscopic observations reveal the presence of Bowen Fluorescence lines and strong high-ionization coronal lines indicating an extreme level of ionization in the system.}
   {The results suggest that the transient can be classified as a Bowen Fluorescence Flare (BFF), a relatively new sub-class of flaring active galactic nuclei (AGN). AT2019aalc can be also classified as an extreme coronal line emitter (ECLE). We found that, in addition to AT2019aalc, another BFF AT2021loi is spatially coincident with a high-energy neutrino event. We propose a repeating TDE scenario within an AGN framework to explain the multi-wavelength properties of AT2019aalc and suggest a possible connection among ECLEs, BFFs, and TDEs occurring in AGNs.}


   \maketitle

\section{Introduction}
Tidal disruption events (TDEs) are rare transient events, which occur when a star approaches a supermassive black hole (SMBH) and the tidal forces of the latter rip the star apart \citep{1988Natur.333..523R}. About half of the star's material is accreted around the black hole, generating a luminous outburst across the electromagnetic spectrum. TDE optical light curves typically exhibit a rapid, colorless rise followed by a fast decay on monthly timescales. Several TDEs were detected in X-rays with soft spectra typically explained by late-time accretion disk formation \citep{2020SSRv..216...85S}. In addition, a handful of TDEs produced detectable radio emission, explained by delayed\footnote{Note that the SMBH accretion rate does not peak on the same timescale as the mass
fall-back rate.} non-relativistic outflows \citep{delay1,delay2}, or only in four instances by newly launched relativistic on-axis jets \citep{2012ApJ...748...36B,2015MNRAS.452.4297B,2012ApJ...753...77C,jetted_tdes,2022cmc}. Spectroscopically, TDEs are characterized by a strong blue continuum and broad emission lines ($> 10^{4}$\,km s$^{-1}$) e.g., H$\alpha$ and H$\beta$. In a few cases, Bowen Fluorescence \citep[BF:][]{1934PASP...46..146B,1935ApJ....81....1B} lines e.g., N\,{\sc iii} at $4640 \AA$ appear \citep[e.g.,][]{2019ApJ...873...92B,spectra2}.

TDEs generally produce one optical flare, however, in a few cases a re-brightening episode was detected, even years after the initial flare. These might be explained in the following way. A star on a grazing orbit might be partially disrupted and only a fraction of the stellar mass will be deposited onto the SMBH. The secondary disruption of a surviving core after it reaches the pericenter again results in re-brightening.  This scenario is naturally explained by the Hills capture mechanism \citep{1988Natur.331..687H}, where a binary star system is disrupted by the SMBH, capturing one star on a highly eccentric orbit prone to repeated partial disruptions. In this process, the captured star’s orbit brings it close enough to the SMBH multiple times, causing episodic stripping of stellar material and producing recurring flares over timescales of months to years. Only a few partial TDEs or candidates are known to date. These are AT2020vdq \citep{2020vdq}, ASASSN$-$14ko \citep{2021ApJ...910..125P,14ko2,14ko3,2023ApJ...956L..46H}, eRASSt J045650.3$-$203750 \citep{2023A&A...669A..75L,2024A&A...683L..13L},  RX J133157.6-324319.7 \citep{2022RAA....22e5004H},  AT2018fyk \citep{fyk1,fyk2,2024ApJ...970..116W,2024ApJ...971L..31P}, AT2019avd \citep{avd1,2021ApJ...920...56F,avd2,avd3,avd4}, AT2019azh \citep{2021MNRAS.500.1673H}, AT2022dbl \citep{2024ApJ...971L..26L,2024arXiv241215326H,2025ApJ...987L..20M} and AT2021aeuk \citep{at2021aeuk_jingbo}.

Based on the observed properties of TDE host galaxies, \cite{2014ApJ...792...53V} estimated a TDE rate of a few $\times10^{-5}$ per year per galaxy, which is well borne out by recent data \citep{2023ApJ...955L...6Y} and earlier observational estimates \citep{1999MNRAS.306...35S}. Importantly, a similar fraction of TDEs should occur in AGN hosts \citep[e.g.,][]{tdes_in_agn, 2024MNRAS.527.8103R}. These transients are challenging to distinguish from regular AGN variability and only a few candidate TDEs occurring in AGN (in the following: TDE-AGN) have been studied earlier (e.g., PS16dtm - a TDE in a narrow-line Seyfert-1 (NLSy1) galaxy \citep{2017ApJ...843..106B} or the multiple soft X-ray flares in IC 3599 \citep{2015A&A...581A..17C} and GSN 069 \citep{2018ApJ...857L..16S}). In contrast with TDEs seen in earlier quiescent galaxies, these transients are characterized by slowly decaying optical emission and bumps tend to appear after the peak. It is worth noting that transients classified as Ambiguous Nuclear Transients (ANTs; e.g., \cite{2025MNRAS.537.2024W}), as well as several of the events studied by \cite{2021ApJ...920...56F}, could also be TDEs in AGN hosts, potentially increasing their overall number.

Recently, \cite{Trakhtenbrot} identified a new class of spectroscopically unique flares from accreting SMBHs. In $2019$, the transient events AT2017bgt, F01004-2237 and OGLE17aaj were identified as the first members of the class based on their unique optical spectroscopic features. Later, the peculiar X-ray transient AT2019avd \citep{avd1,2021ApJ...920...56F,avd2,avd3,avd4} was found with remarkable common properties, however, its classification as an AT2017bgt-like transient is unclear. A more promising case is AT2021loi \citep{AT2021loi}, a UV-bright transient event with very similar multi-wavelength properties to the three flares studied by \cite{Trakhtenbrot}. These events show unobscured AGN-like optical spectra with significant and persistent BF lines, high UV-to-optical ratio and optical emission that decays on yearly timescales. These AGN are characterized by accretion that intensifies significantly over long timescales. Hereafter, we will refer to this new class of SMBH flares as Bowen Fluorescence Flares (BFFs), following \cite{AT2021loi}. Using the Zwicky Transient Facility \citep[ZTF,][]{2019PASP..131a8003M} Public Survey data, \cite{2023ApJ...957...57D} found only one BFF candidate (AT2021seu) among $223$ spectroscopically classified transients, suggesting that these flares are rare. Two other candidates, namely AT2022fpx, \citep{2024MNRAS.532..112K} and AT2020afhd \citep{2024TNSAN..53....1A} were found later.

In recent years, two nuclear transients were found to be spatially coincident with neutrino events detected by the IceCube Neutrino Observatory based on ZTF follow-up campaigns of IceCube neutrino alerts \citep{2023MNRAS.521.5046S}. The TDE AT2019dsg is likely associated with the  neutrino, IC191001A \citep{Bran} and a further promising association, between the candidate TDE AT2019fdr and  IC200530A \citep{Tywin} points to TDEs as a potential new class of sources of the extragalactic neutrinos. Both sources were identified as having luminous mid-infrared emission, and a search for similar flares \citep{Lancel} led to the identification of the transient event AT2019aalc (a.k.a. ZTF19aaejtoy) with the neutrino alert IC191119A. In addition, two candidate obscured TDEs \citep{hidden} and the  extremely energetic candidate TDE AT2021lwx \citep{lwx} were associated with high-energy neutrinos. A recently claimed association between a neutrino flare detected by the IceCube Observatory and an X-ray flare from the TDE ATLAS17jrp suggests that the X-ray emission of these sources plays a crucial rule in producing neutrinos \citep{xray_neu}.

In this paper, we focus on the nuclear transient event and neutrino source candidate AT2019aalc, especially on the significant optical re-brightening that started four years after its initial flare \citep{re-br}.

The paper is organized as follows. We discuss the discovery of the transient event and give an insight about its host galaxy in Sect.~\ref{sect_2},  summarize our observing campaign and the data reduction in Sect.~\ref{sect_3}, present the results of the observations in Sect.~\ref{sect_4}, discuss them in Sect.~\ref{sect_5} and finally give a summary and conclude in Sect.~\ref{sect_6}. Throughout the paper, we adopt a flat $\Lambda$CDM cosmological model with parameters $H_0=70\,\textrm{km\,s}^{-1}\textrm{\,Mpc}^{-1}$, $\Omega_\Lambda=0.73$, and $\Omega_\textrm{m}=0.27$.  In this model \citep{2006PASP..118.1711W}, $1$\,mas angular size corresponds to $0.72$\,pc projected linear size at the source redshift $z = 0.0356$ \citep{2012ApJS..203...21A}. This redshift corresponds to a luminosity distance of $158.3$\,Mpc. All magnitudes are given in AB system \citep{1974ApJS...27...21O} unless stated otherwise.

\section{The transient event: AT2019aalc}
\label{sect_2}

An analysis of the infrared properties of AT2019dsg and AT2019fdr revealed a strong reverberation signal, particularly an the infrared dust echo, stemming from reprocessed emission of optical/UV photons to the infrared by interaction with hot dust around the SMBH, at a distance of $0.1-1$\,pc \citep{2016ApJ...828L..14J,2016ApJ...832..188D, 2016MNRAS.458..575L, 2021SSRv..217...63V,2021ApJ...911...31J,  Lancel}. In addition to the exceptionally high infrared luminosities  of these transients, the neutrino detection times appear to be temporally consistent with the infrared luminosity peaks \citep{Bran,Tywin}.
This discovery motivated a systematic search for neutrino emission using an extended sample of black hole flares \citep{Lancel}. This archival search revealed the nuclear transient AT2019aalc, with a powerful optical flare and dust echo emission in spatial coincidence with the high-energy neutrino event, IC191119A \citep{2019GCN.26258....1I}, detected by the IceCube Observatory.

\begin{figure}[H]
  \centering
  {\includegraphics[width=95mm]{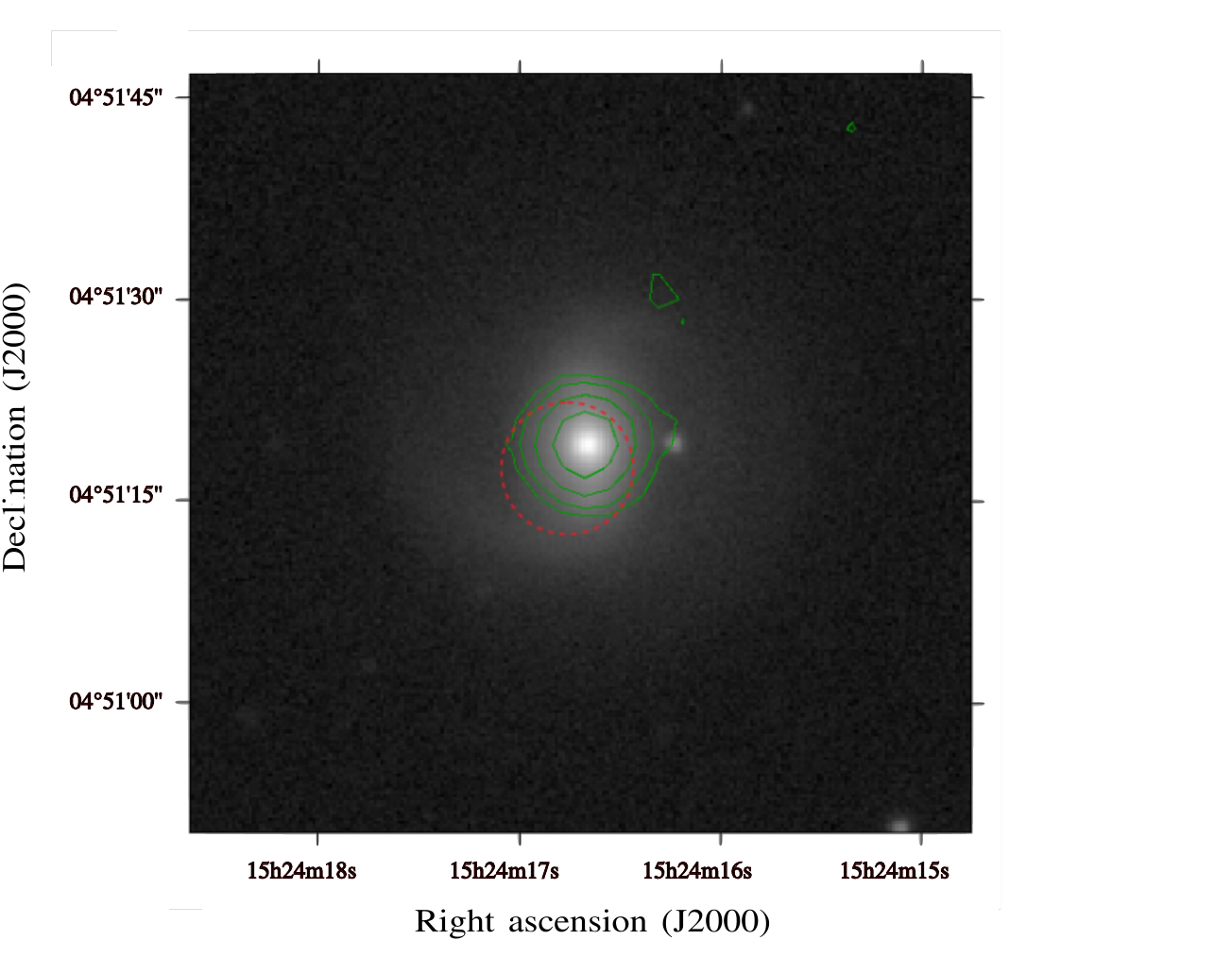}}
  \caption{Pan-STARRS i-band pre-flare image centered on the host galaxy position of the transient event AT2019aalc. The object located to the west of the galaxy core is a foreground star. The green contours represent the pre-flare FIRST radio observations taken at $1.4$\,GHz. The lowest contour level is drawn at $3\sigma$ image noise level corresponding to $0.4\textrm{\,mJy\,beam}^{-1}$. Additional positive contour levels increase by a factor of two. The peak brightness is $5.8\textrm{\,mJy\,beam}^{-1}$. The red circle points to the \textit{Swift}-XRT $0.3-10$\,keV astrometrically-corrected position of AT2019aalc, determined using the \textit{Swift}-XRT data products generator \citep{astro_swift}. Its size indicates the $90\%$ confidence positional error.}
  \label{pic:overlay}
\end{figure}

\begin{figure}
  \centering
  {\includegraphics[width=90mm]{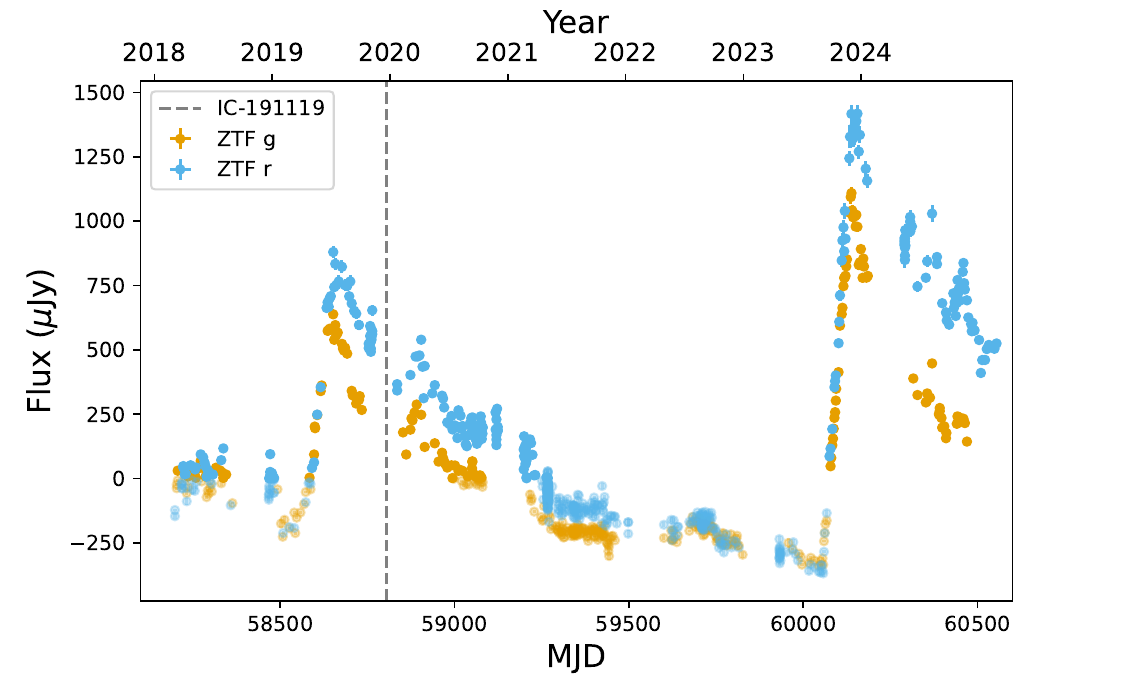}}
  \caption{Optical light curve of AT2019aalc derived from ZTF difference photometry. The transparent markers indicate negative flux i.e. the flux decreased below the mean flux of the reference images. Two distinct flares can be seen. The second flare peaked four years after the first one.  The blue vertical line indicates the arrival time of the neutrino event IC191119A associated with the transient.}
  \label{pic:ztf_diff}
\end{figure}

The host galaxy of AT2019aalc (2MASX J15241665+0451192) is a barred spiral (SBa) galaxy hosting an active nucleus \citep{{2013AJ....146..151O}}. Based on its Sloan Digital Sky Survey \citep[SDSS:][]{sdssdr7} spectrum it was further classified as a broad-line Seyfert-1 galaxy \citep{2013ApJ...763..133T,2019ApJS..243...21L}. The galaxy was detected in archival radio sky surveys e.g., the Very Large Array Faint Images of the Radio Sky at Twenty-cm \citep[VLA FIRST:][]{1998AJ....115.1693C}. Fig.~\ref{pic:overlay} shows the Pan-STARRS \citep{2016arXiv161205560C} $i$-band optical image of the galaxy, together with the contours of the FIRST radio data (at $1.4$\,GHz) and the position of the X-ray counterpart
detected by the \textit{Swift}/XRT in $2023/2024$. The galaxy is not present in any X-ray catalogs prior to the optical discovery of the transient in $2019$. In addition to the  significant IR echo, AT2019aalc shares multi-wavelength characteristics with the two other neutrino candidate transients: a soft X-ray spectrum and transient radio emission \citep{Lancel}. Moreover, similarly to the other two instances, the infrared peak of the transient event is temporarily coincident with the detection time of the likely associated neutrino event. Notably, \cite{2023ApJ...948...42W} estimated the highest neutrino fluence for AT2019aalc among the studied candidates, due to its high estimated SMBH mass ($M_\textrm{BH} = 10^{7.2} M_{\odot}$, \citealp{Lancel}) and low redshift. Since the proposed association with the high-energy neutrino event was announced roughly two years after the discovery of the transient event, the source was not at the focus of attention. Thus, the lack of spectroscopic monitoring during the first optical flare hindered a clear classification of the event. Nevertheless, out of the more than $10^{4}$ AGN detected by the ZTF, less than $1\%$ show similarly rapid and large outbursts \citep{Tywin,Lancel}, suggesting a very unusual transient. Interestingly, roughly four years after the first flare, a significant optical re-brightening started in mid-May 2023 \citep{re-br}, illustrated by the long-term ZTF difference light curve shown in Fig.~\ref{pic:ztf_diff}.

\section{Observations and data reduction}
\label{sect_3}
\subsection{Optical/UV}
\subsubsection{ZTF}
ZTF forced photometry at the location of AT2019aalc was obtained following the recommendations outlined in \cite{2019PASP..131a8003M}. We corrected for extinction by adopting a galactic extinction of $E(B-V) = 0.039$ \citep{2011ApJ...737..103S}. As shown in Fig.~\ref{pic:ztf_diff}, the forced photometry yields a negative flux in the period between the two flares (and also just before the first flare). This implies that during these epochs, the flux decreased below the mean flux of the reference images. For the $r$-band, these were obtained between March 2018 and May 2018. While the baseline used to obtain the difference flux of the $g$-band is based on images obtained between March and September, 2018.

\subsubsection{UVOT}
The \textit{Neil Gehrels Swift Observatory} \citep{2004ApJ...611.1005G} performed $33$ Target of Opportunity (ToO) observations (project codes: 18972, 19188, 19993, 20997 and 21692, Proposers: Reusch, Veres \& Sniegowska) on AT2019aalc from 2023-06-23 to 2025-01-02 with a total exposure time of $\approx55$\,ks. The Ultraviolet and Optical Telescope (UVOT, \citealt{uvot}) onboard the satellite observed the target field with the following four filters in each epoch: UVW2 (central wavelength, 1928 $\AA$), UVM2 (2246 $\AA$), UVW1 (2600 $\AA$) and U (3465 $\AA$). We analyzed the UVOT images using the \textit{Swift}/UVOT tools included in the \textsc{heasoft 6.30} software package. We measured the flux using \textsc{uvotsource}, applying a circular aperture of $15$\,arcseconds (which captures nearly all the flux of the host galaxy). The background was measured using four 6\,arcsecond regions in four quadrants away from the target source. To estimate the UV flux before the flare, we use GALEX \citep{Martin05} observations obtained in 2007 and 2011. We use the same 15" circular aperture that was applied to the {\it Swift}/UVOT observations and use the \textsc{gPhoton} software \citep{Million16} to extract the flux. Over the four year baseline of the GALEX observations, the NUV magnitudes change by about 0.3~mag. Following \citet{vanVelzen20}, we combine the mean GALEX flux with the SDSS \citep{york02} magnitudes and apply Prospector \citep{Johnson17} to find the best-fit stellar population synthesis model \citep{Conroy09} that describes these data. From this model we estimate the baseline flux in the UVOT filters and this baseline is subtracted to obtain the difference flux. The baseline flux in the UVM2 band is 17.2 mag, implying an increase of 1.8~mag relative to the peak of the UVM2 light curve. This large increase implies that, at the peak of the flare, we are not sensitive to the value of the baseline flux. This is useful because the baseline of the ZTF forced photometry was obtained around 2018 (at the start of the ZTF survey), while the baseline that is used to obtain the UVOT difference flux is mainly determined by the GALEX observations that were obtained a decade earlier.

\subsection{Infrared}
The Palomar 200-inch (P200) near-infrared telescope equipped with the Wide Field Infrared Camera (WIRC, \citealp{2003SPIE.4841..451W}) observed AT2019aalc on 2023-07-05. The imaging was obtained in the J-, H,- and Ks-bands. The data was processed with the \textsc{Swarp} software \citep{2002ASPC..281..228B}. We performed background subtraction using the \textsc{photutils.background} class and measured the flux using a circular aperture with a radius of $9$\,pixels. The corresponding zero-point magnitudes for this aperture size were calculated using the automatic pipeline for each filter. The error bars are the computed Poisson and background error of the photometry\footnote{\url{https://www.stsci.edu/hst/instrumentation/wfc3/data-analysis/photometric-calibration/ir-photometric-calibration}} and we also took into account the uncertainty of the derived zero-point magnitudes.

We obtained mid-infrared photometry from archival observations of the Wide-Field Survey Explorer \citep[WISE,][]{2010AJ....140.1868W}, which was continuously monitoring the sky at $3.4\,\mathrm{\mu m}$ (W1) and, due to the NEOWISE project \citep{2011ApJ...731...53M}, at $4.6\,\mathrm{\mu m}$ (W2). NEOWISE visits any point in the sky roughly every six months, with each visit consisting of about ten exposures. Using \texttt{timewise} \citep{neckerJannisNeTimewiseV02024, 2025A&A...695A.228N}, we downloaded all available multi-epoch photometry data from the AllWISE data release and single-epoch photometry from the NEOWISE-R data releases and stacked the data per visit to obtain robust measurements. All selected single-exposure photometry data points are within $1$\,arcsecond of the nucleus of the host. The resulting light curve spans almost $14$\,years, starting in February $2010$ until July 2024.

\subsection{X-ray}
\label{subsect:x_data}
The X-ray Telescope \citep[XRT,][]{2005SSRv..120..165B} observations, in the energy range of $0.3-10$\,keV, were obtained simultaneously with the UVOT observations. The observations were performed in photon counting mode. We performed the data reduction using the Space Science Data Center (SSDC) interactive archive in the following way: first, cleaned event-files were produced using the \textsc{xrtpipeline} and \textsc{xrtexpomap} tasks. The \textsc{ximage} tool allowed us to determine the source detection level and its position. In case of non-detections, $3\sigma$ upper limits for the count rate were derived and converted to flux using WebPIMMS\footnote{\url{https://heasarc.gsfc.nasa.gov/cgi-bin/Tools/w3pimms/w3pimms.pl}} and assuming the mean power-law index of the observations that resulted in detections. When our target source was detected above the $3\sigma$ level, we ran \textsc{swxrtdas} in order to extract the source and the background spectra. Source counts were retrieved from a circle with a radius of $20$ pixel ($47$\,arcseconds), while background counts were extracted using an annular region with $80/120$ pixel inner/outer radius. Finally, the \textsc{grppha} tool of the \textsc{heasoft 6.32} software package was used to perform background subtraction using the source and background spectra.  We binned each spectrum to have a minimum of $3$ counts per spectral bin. 

We fit the individual X-ray spectra using the XSPEC fitting environment \citep{swift}.  First, we have determined the  Galactic neutral hydrogen column density, $N_{\textrm{H}}$, based on the \textsc{HI4PI} survey \citep{2016A&A...594A.116H}.  The spectrum was then fitted using Cash-statistic (C-stat, \citealt{1976A&A....52..307C}). To find the best-fitting model, we first fitted the stacked spectrum with a absorbed power-law (\textsc{Tbabs*powerlaw} in XSPEC) model, as  the last XMM data of AT2019aalc before our monitoring program was best-fit with a power-law model \citep{2023ATel16118....1P}. When fitting the model, the galactic column density was fixed whilst the power-law index and the normalization were free to vary. The resultant fit quality is C-Stat $= 183.3$ for $165$ degrees of freedom. We then added a blackbody model component, to take into account a possible soft excess. The blackbody temperature and the normalization of the new component were also free to vary. We fit again, which resulted in a better fit with C-Stat $=170.5$ for $163$ degrees of freedom. We implemented the likelihood ratio test (lrt in XSPEC) to estimate the statistical significance of the inclusion of the blackbody component. After running $100$ simulations, the test resulted only in smaller likelihood differences than the observed one, clearly implying that the difference is unlikely to be due to random fluctuations. Therefore, accounting for the soft X-ray excess results in a significantly better fit. 

\subsection{Radio}
\subsubsection{VLBI observations}
We observed AT2019aalc with the European VLBI Network (EVN) and the enhanced Multi-Element Remotely Linked Interferometer Network \citep[e-MERLIN,][]{2004SPIE.5489..332G}.  These observations were carried out at $1.7$\,GHz (L-band) on 2023 June 09 (project code: EV027, PI: Veres).  The  observation time was $10$\,hr. The experiment was performed in phase-referencing mode with $5$-min duty cycles (with spending $3.5$\,min on the target during each cycle) which resulted in an on-source time of $6.7$\,h including fringe-finder scans and slewing. The data were recorded in left and right circular polarizations with a data rate of $1$~Gbit\,s$^{-1}$. The correlation was done at the EVN Data Processor \citep{JIVEcorrelator} at the Joint Institute for VLBI European Research Infrastructure Consortium (Dwingeloo, The Netherlands) with an averaging time of $2$\,s. The Compact Symmetric Object (CSO) 1521$+$0430 (a.k.a. 4C 04.51) at an angular distance of $0.83\degr$ was observed as phase-reference calibrator.

The data were calibrated with the U.S. National Radio Astronomy Observatory Astronomical Image Processing System (AIPS, \citealt{aips}) software package. The standard procedures including time flagging, amplitude calibration, correction for the dispersive ionospheric delays and global fringe-fitting of the calibrator sources were followed. We imaged the phase-calibrator with the \textsc{Difmap} software \citep{difmap}, using the hybrid mapping technique \citep{hybridmapping}. Using the \textsc{clean} component (CC) model we built up in \textsc{Difmap}, we repeated the fringe-fitting in \textsc{AIPS} for the phase-reference calibrator source to correct its structure. Finally, we interpolated the solutions of the last fringe-fitting to the data of our target source. We imaged the target source using the hybrid mapping technique in \textsc{Difmap}. At the end of the procedure, several steps of amplitude and phase self-calibration were performed. A simple elliptical Gaussian modelfit component \citep{1995ASPC...82..267P} was fitted directly to the visibility data. Finally, to decrease the noise level, we performed $1000$ \textsc{clean} iterations in the residual image with a small loop gain of $0.01$.

\subsubsection{ATCA monitoring}
The Australia Telescope Compact Array (ATCA)\footnote{\url{https://csiropedia.csiro.au/australia-telescope-compact-array/}}, located at Narrabri, New South Wales, Australia, was used to monitor AT2019aalc. The observations were conducted between 2023-06-28 and 2024-05-04, utilizing the array in the 6km (6A and 6D) and (H168) configuration. The observations were conducted at frequencies of $2.1$\,GHz, $5.5$\,GHz, $9$\,GHz, $17$\,GHz and $19$\,GHz. The flux calibrator was the standard ATCA primary calibrator, 1934–638, and 1548+056 was used as phase calibrator. The observations are summarized in Table~\ref{table:atca}.

The data reduction procedures were based on the Multichannel Image Reconstruction, Imaging Analysis and Display \citep[\texttt{miriad}:][]{miriad}. The data reduction procedure was conducted in accordance with standard protocols. In order to mitigate the impact of radio frequency interference (RFI) during flux and phase calibration, the interactive flagging tool \texttt{bflag} was utilised. Furthermore, automated flagging routines \texttt{pgflag} were applied to address interference for the source. Any corrupted data identified during calibration was manually flagged.

Due to the limited ($u,v$)-coverage and the low elevation of the source, which resulted in a very elongated beamshape, we were not able to perform self-calibration and cleaning. Fortunately, AT2019aalc always dominated the flux within the resolution elements of our short observations. Therefore, we imaged the region around our source for each observation using an oversampling of the synthesised beam by a factor of at least 10. Under the assumption that AT2019aalc is always a point-source, we can take the highest pixel value in the image as our measured flux. Due to the uncertainties this method introduced, we used a plain $10$\,\% flux error. 

\begin{table*}
\centering
\caption{Summary of the ATCA radio continuum emission observations of AT2019aalc}
\begin{tabular}{l|cccc}
\toprule
Date & Exp. time / min & Configuration & Central Freq. / GHz & Bandwidth / GHz \\
\hline
28.06.2023 & 330 / 180 & 6D & 2.1 / 7.0 & 1 / 4 \\
30.06.2023 & 90 / 180 & 6D & 2.1 / 7.0 & 1 / 4 \\
08.07.2023 & 120 / 120 & 6D & 2.1 / 7.0 & 1 / 4 \\
19.07.2023 & 210 & 6D & 2.1 & 1\\
09.08.2023 & 120 / 150 & 6D & 2.1 / 7.0 & 1 / 4 \\
19.10.2023 & 180 / 180 & H168 & 2.1 / 7.0 & 1 / 4 \\
09.02.2024 & 60 / 60 / 60 & 6A & 2.1 / 7.0 & 1 / 4 / 4\\
29.03.2024 & 60 / 60 / 60 & 6A & 2.1 / 7.0 / 17.0 & 1 / 4 / 4 \\
\bottomrule
\label{table:atca}
\end{tabular}
\end{table*}

\subsection{Spectroscopic observations}
\label{subsect:spectro_data}
We observed the spectrum of AT2019aalc on 2021-07-06 using the Low Resolution Imaging Spectrograph \citep[LRIS:][]{1995PASP..107..375O} mounted on the Keck-I telescope at Maunakea (PI: Kulkarni).  Later, during the second optical flare, we obtained five more optical spectra of AT2019aalc using the following instruments: LRIS (PI: Kulkarni), DeVeny Spectrograph mounted on the Lowell Discovery Telescope (LDT) in Arizona (PI: Hammerstein) and the Double Beam Spectrograph \citep[DBSP,][]{1982PASP...94..586O} mounted on the $200$ inch telescope at Palomar Observatory, California (PIs: Kulkarni and Kasliwal). The spectra were reduced following the standard procedures using the automated reduction pipeline of the LRIS \citep{2019PASP..131h4503P}, the software package \textsc{PypeIt} \citep{2020JOSS....5.2308P} for DeVeny, and the software packages \textsc{DBSP DRT} \citep{2022JOSS....7.3612M} and \textsc{PypeIt} \citep{2020JOSS....5.2308P} for the DBSP.  Details of these observations of AT2019aalc are summarized in Table~\ref{table:sp} while the spectra are shown in Fig.~\ref{fig:a1}.

To determine the type of AGN prior to the flares, we performed a stellar population synthesis fit of the pre-flare SDSS spectrum with \citep[\textsc{pPXF}:][]{2004PASP..116..138C,2017MNRAS.466..798C,2023MNRAS.526.3273C} and the newest \textsc{GALAXEV} \citep{2003MNRAS.344.1000B} models (\textsc{CB2019}), assuming a \cite{2001MNRAS.322..231K} initial mass function (IMF) and an upper mass limit of $100$, as templates (see Fig.~\ref{fig:a2}). Beforehand, the spectrum was corrected for Milky Way extinction using the dust reddening maps by \cite{2011ApJ...737..103S} and the \cite{1989ApJ...345..245C} extinction curve. Furthermore, it was redshift corrected and logarithmically rebinned. We included several emission lines in the fit and assumed four kinematic components: one for all stellar templates, one for all forbidden lines, one for all allowed emission lines, and a last one for a second, broad component of all Balmer lines up to $\mathrm{H}\delta$.  Additionally, we included two separate dust components, each one with the extinction curve of \cite{1989ApJ...345..245C}: a first one for stellar, and a second one for nebular emission. We therefore fixed the Balmer line ratio and limited doublets of the preinstalled lines to their theoretical ratio using the keywords \textsc{tie\_balmer} and \textsc{limit\_doublets} in \textsc{pPXF}. However, we only fixed the ratio for the strongest lines of the Balmer series up to $\mathrm{H}\delta$ to avoid possible contamination of $\mathrm{H}\epsilon$ by the [Ne\,{\sc iii}]$ \lambda 3967$ line. For the fit with \textsc{pPXF} we included additive and multiplicative polynomials up to tenth order (\textsc{degree=10}, \textsc{mdegree=10}) to model the AGN continuum. In order to correct for intrinsic dispersion caused by the spectrograph, we assume a resolving power of $1500$ at $3800\,\text{\AA}$ and $2500$ at $9000\,\text{\AA}$.\footnote{\url{https://www.sdss4.org/dr17/spectro/spectro_basics/}} Linear interpolation is used to calculate values for wavelengths in between. 

The newly observed spectra were fitted using the galaxy spectrum fitting tool \textsc{PyQSOFit} \citep{qso1,qso2,qso3}. The continuum of each spectrum was modeled with a combination of the following components: a power-law to represent the AGN continuum, polynomials to capture any residual smooth variations, and an Fe II template for the iron emission complex. Emission lines in the H$\beta$ and H$\alpha$ complexes were modeled using a combination of broad and narrow Gaussian profiles to represent different kinematic components. The high-ionization coronal lines of interest were fitted with additional Gaussian profiles. The H$\beta$ complex shows a strong double-peaked feature around $4660 \AA$. This region is known as Bowen Fluorescence complex. It was fitted with a set of broad Gaussian components. We computed parameter uncertainties using MC resampling. Examples of the fitted spectra and the inferred emission line parameters are given in Fig. \ref{fig:a3} and in Table \ref{table:a2}, respectively.

\section{Results}
\label{sect_4}
\begin{figure*}
  \centering
  {\includegraphics[width=\textwidth]{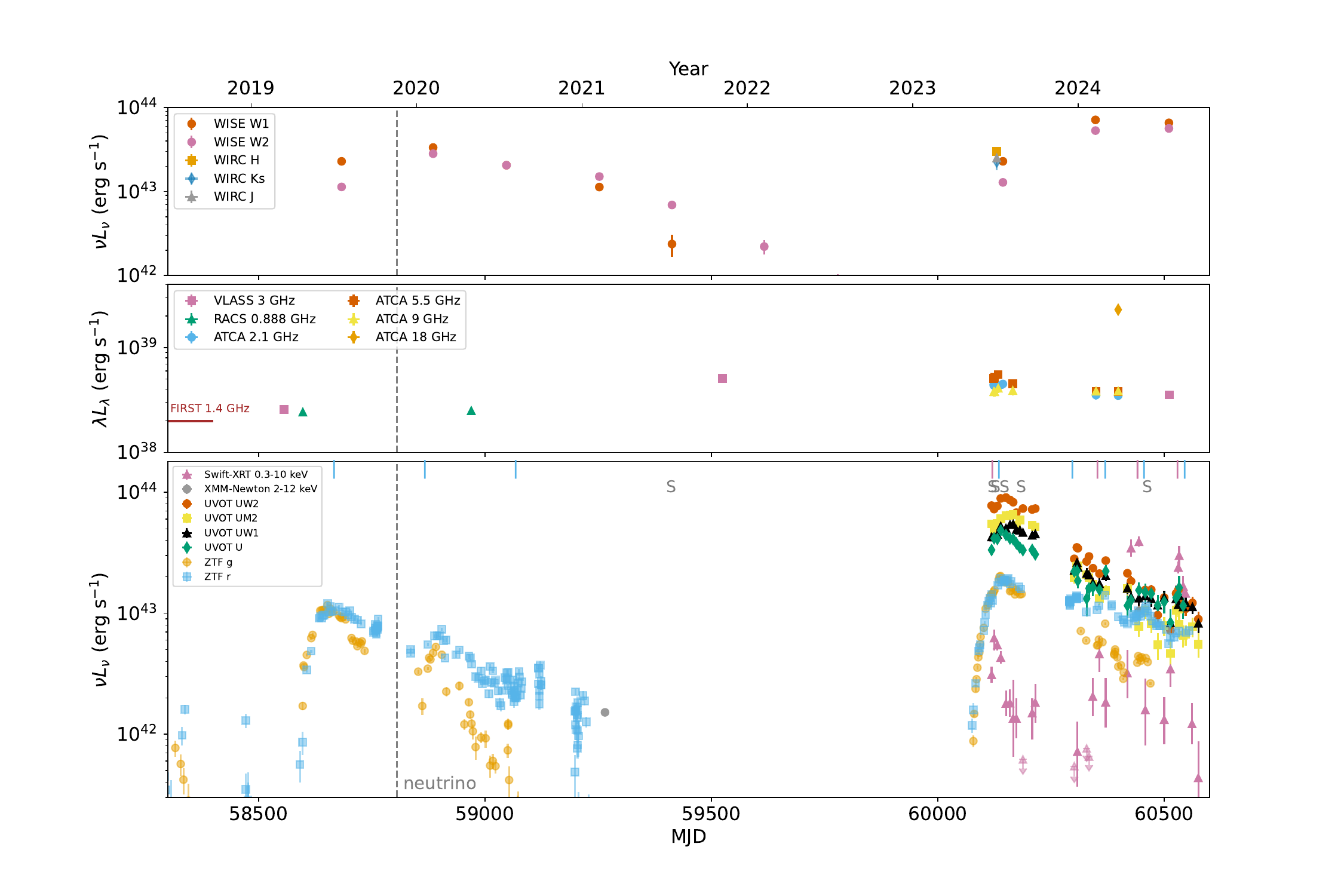}}
  \caption{Multi-wavelength (bottom plot: X-ray and extinction-corrected and host-subtracted optical/UV, middle plot: radio and upper plot: extinction-corrected and host-subtracted IR) light curves of AT2019aalc. Here we show only the positive ZTF fluxes based on the forced photometry results. The dashed gray vertical line indicates the IceCube detection of the high-energy neutrino event IC191119A. The short, purple vertical lines on the top of the bottom panel indicate the peak times of the X-ray flares while the blue ones show the times the main optical $r$-band flare and later the bumps. The grey 'S' letters represent the observing times of the optical spectra..}
  \label{pic:lcs}
\end{figure*}

\subsection{Optical/UV}
The optical and UV light curves of AT2019aalc are shown in the bottom panel of Fig.~\ref{pic:lcs}. The transient was first detected by the ZTF in January $2019$ with an $r$-band magnitude of $18.2$\,mag. AT2019aalc brightened to a peak magnitude of $r \approx 16.7$ ($M = -19.4$) reached on 2019-06-18 on a timescale of $\approx 60$\,days. The second flare started mid-May $2023$, approximately four years after the initial flare \citep{re-br}. The second flare evolved very similarly to the original one. From the beginning of the second flare, a $2.8$\,mag rise was seen up to the peak of the flare on 2023-07-13 with $r \approx 16.2$\,mag ($M = -19.8$), on a timescale of $\approx 60$\,days.  The second flare peaked at roughly $1.5$\,times the peak luminosity of the first flare. The peak of the second flare was followed by a short plateau phase that lasted for around $25$\,days. Both flares declined very slowly, on timescales of years. The first flare decayed with a power-law index of $b \approx -0.45$  while the second with  $b \approx -0.23$ in the $r$-band. The decaying phases are not monotonic, as bumps tend to appear. During the second optical flare, these bumps appear four times: end of December $2023$, mid-March $2024$, end of May $2024$ and mid-August $2024$ (see Figure \ref{pic:lcs}). Both main optical flares brightened with constant color ($g-r \approx 0$) and showed reddening when decaying. 

\begin{figure}
  \centering
  {\includegraphics[width=95mm]{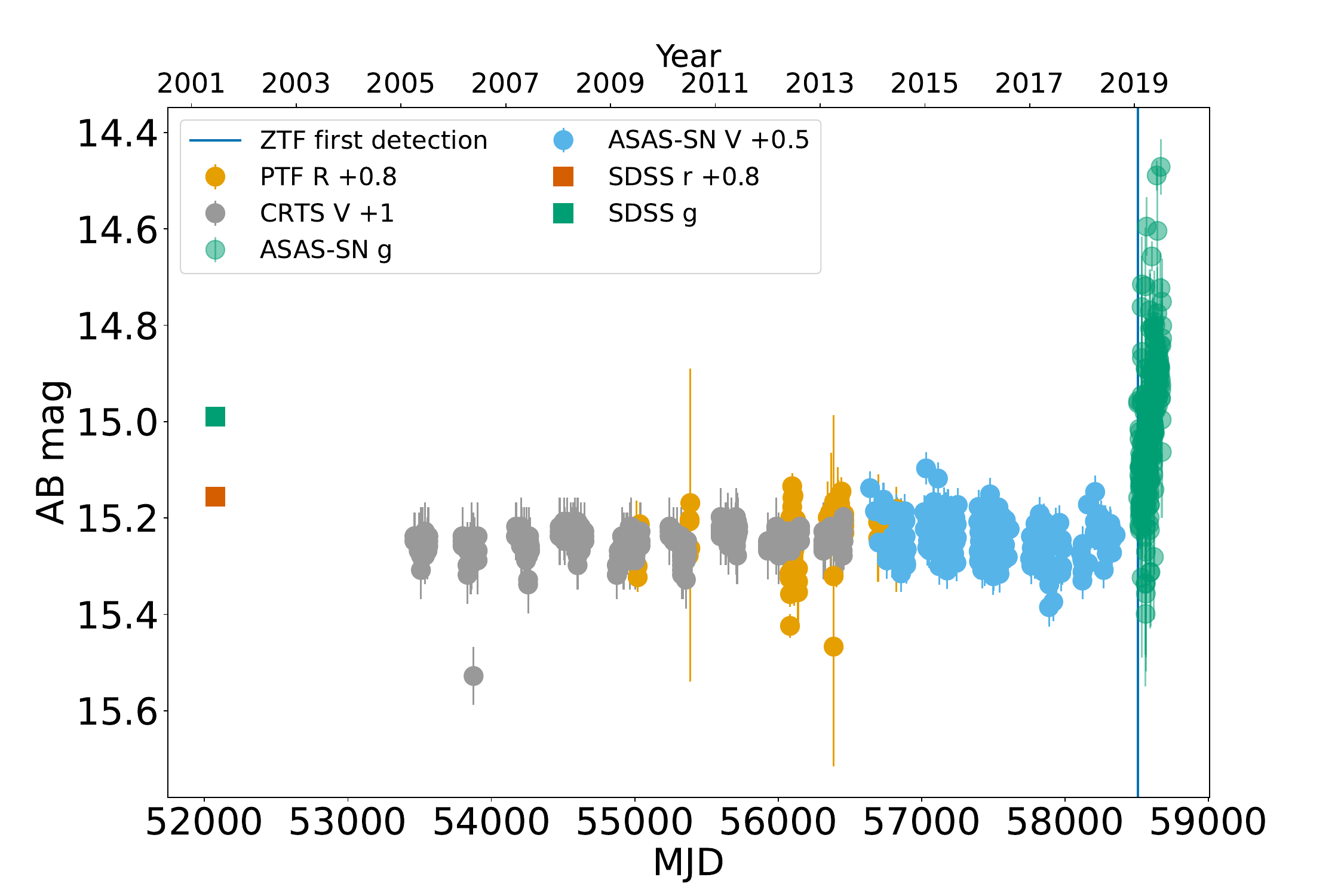}}
  \caption{Pre-flare long-term optical light curves of the AGN hosting AT2019aalc, starting approximately $14$\,years before the discovery of the transient in January $2019$, which is indicated with a blue vertical line.}
  \label{pic:opt_var}
\end{figure}

Prior to the discovery of AT2019aalc, the host galaxy of the transient was monitored by several optical surveys. To study the pre-discovery optical variability of AT2019aalc, we performed aperture photometry of the All-Sky Automated Survey for Supernovae \citep[ASAS-SN:][]{2014ApJ...788...48S, 2017PASP..129j4502K} and the Catalina Real-Time Transient Survey \citep[CRTS:][]{2011arXiv1102.5004D}.  In addition, Palomar Transient Factory \citep[PTF][]{2009PASP..121.1395L} magnitudes were extracted from the PTF Lightcurve Table\footnote{\hyperlink{https://irsa.ipac.caltech.edu/workspace/TMP_ZIQt0M_4155/Gator/irsa/5627/tbview.html}{https://irsa.ipac.caltech.edu}}. No obvious signs of flaring activity is visible between April $2005$ and the discovery of the transient in January $2019$ by ZTF (see Fig.~\ref{pic:opt_var}).

 The UV monitoring of the source started on 2023-06-23 around $20$\,days before the optical peak of the second flare.  The U-band peak is coincident with the optical peak, while the shorter wavelength bands peaked $2-4$\,weeks later. Furthermore, the U- and UVW2-band peaks occur closer in time than the UVW1 and UVM2 peaks, suggesting a complex SED evolution near the maximum. We note that right after the $r$-band peak, a short plateau phase started in optical and UV. The transient is most luminous at the shortest wavelength ranges covered by UVOT. The UV emission peaks around $L_{uv}\approx10^{44}$\,erg/s in the filter UVW1. After the peak, the UVW2$-$U color shows reddening. Towards the end of the UVOT monitoring, the measured UVM2 magnitudes decreased back to the level of the baseline GALEX-NUV magnitudes measured between $2007$ and $2011$.

Comparing the peak UVM2 magnitude of the UVOT monitoring to the GALEX-DR5 \citep{2011Ap&SS.335..161B} NUV magnitude implies that the UV luminosity increased by a factor of $\approx 7$ with respect to the archival flux. The amplitude of this flare is higher than typically seen for AGN. Only $8$ out of the $305$ AGN presented in the GALEX Time Domain Survey \citep{Gezari13} have larger amplitude NUV variability.

\subsection{Infrared}
\label{subsect:ir_res}
We detected AT2019aalc in all of the three bands observed with the WIRC. The observed magnitudes are plotted in the top panel of Fig.~\ref{pic:lcs}. We compare these to pre-flare magnitudes presented in the 2MASS All-Sky Catalog of Point Sources \citep{2003tmc..book.....C}. The observed magnitudes in April $2000$ are higher with $\approx 0.5$\,mag in J-, and Ks-bands and $\approx 0.6$\,mag in H-band indicating the clear brightening of the IR counterpart with respect to archival data.

\begin{figure}
  \centering
  {\includegraphics[width=70mm]{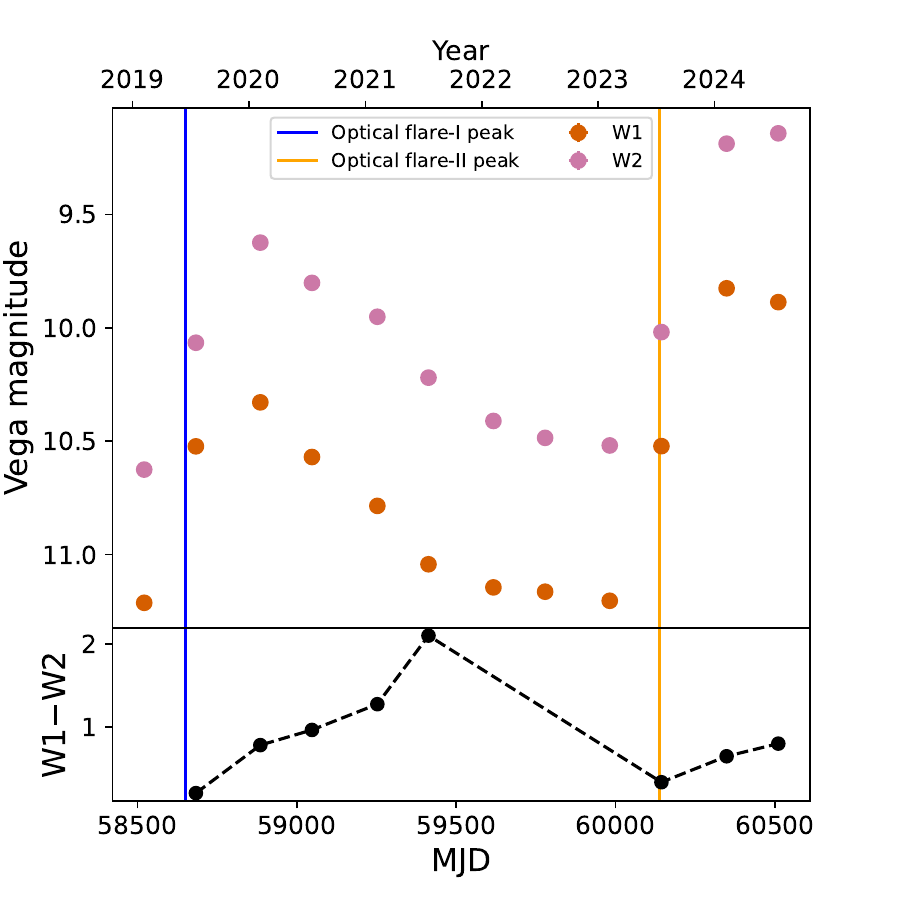}}
  \caption{The long-term IR variability of AT2019aalc. The W1$-$W2 color indicates heated dust as the dominant origin of the IR emission. The second optical flare is accompanied by another IR flaring episode with color change towards the blue. The plotted magnitudes are in the Vega system and have been corrected for extinction but not for host-galaxy contribution, allowing the full evolution between the flares to be visible.}
  \label{pic:wise_var}
\end{figure}

The top panel of Fig.~\ref{pic:lcs} shows the light curve starting just before the first optical flare. The distinct first dust echo flare peaks around six months after the first optical peak. The rise of the dust echo of the second flare also becomes visible in the latest WISE data releases. The light curve varies by around $0.3$\,mag in W1 and $0.4$\,mag in W2 before the dust echo flare, which is consistent with statistical AGN variability \citep{berkEnsemblePhotometricVariability2004}. In contrast, the dust echo presents an increase of $1.0$\,mag in W1 and $1.2$\,mag in W2 after the first optical flare. Following the second optical flare, the IR light curve shows a brightening of around $1.3$\,mag in both bands and is increasing when the optical flare is already decaying. Because the brightness is significantly above the limiting magnitude where Eddington bias becomes an issue for stacked single-exposure photometry \citep{2025A&A...695A.228N}, we can estimate the pre-flare mid-IR color using the median of the light curve prior to the dust echo flare with $m_\mathrm{W1} - m_\mathrm{W2} \approx 0.6$. Although there is clear AGN activity prior to the dust echo flare, this is still below the widely used AGN identification cut of $m_\mathrm{W1} - m_\mathrm{W2} \geq 0.8$ \citep{2012ApJ...753...30S}. However, according to the reliability of WISE selection as a function of a simple W1$-$W2 color selection \citep{2012ApJ...753...30S}, the color of AT2019aalc suggests a reliability of only $\approx 70$\%. With the onset of the dust echo, the IR emission becomes dominated by the heated dust. The change from bluer to redder colors comes from cooling of the dust. The re-brightening of the second dust echo is consequently accompanied by another color change towards the blue, shown in Fig.~\ref{pic:wise_var}. 

Notably, AT2019aalc is not part of the Flaires sample \citep{2025A&A...695A.228N}, a list of dust-echo-like infrared flares interpreted as extreme AGN accretion events or TDEs. This is because the broad application of the Flaires pipeline necessarily included a strict cut on extraneous variability that AT2019aalc did not pass. The authors did note however that a desirable relaxation of this criterion would include AT2019aalc. The source is, however, part of the dust echo sample of \cite{Lancel} and, notably has the highest dust echo flux of all ZTF transients.

\subsection{X-ray}
\label{subsect:x_results}
\begin{figure}
  \centering
  {\includegraphics[width=90mm]{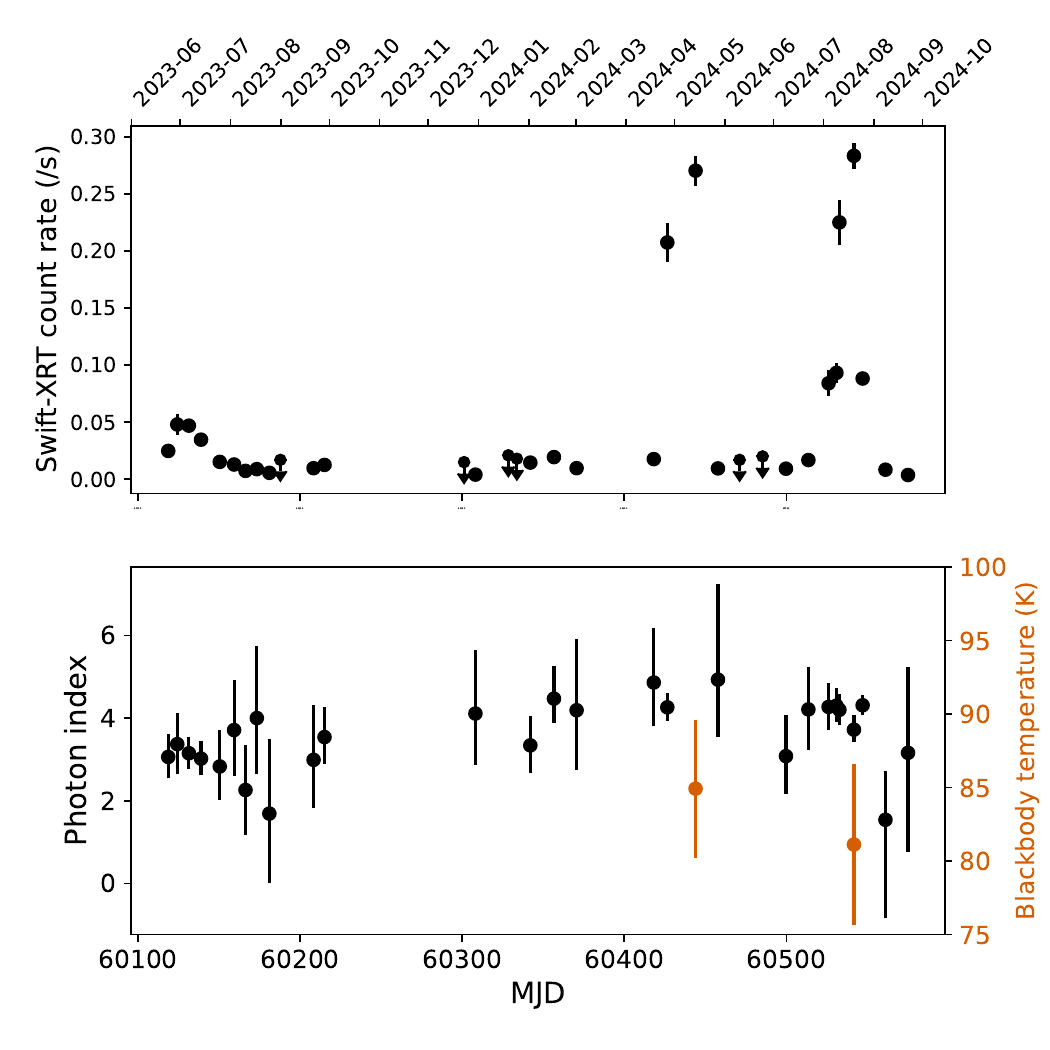}}
  \caption{The \textit{Swift}-XRT count rate and photon index variability of AT2019aalc during the observing campaign. In most cases the data is well described by a power-law spectral model (black), except during the two recent flaring episodes, where blackbody or combined models are preferred (red). The blackbody temperature is indicated on the red y-axis.}
  \label{pic:countrate_and_ph}
\end{figure}

The $0.3-10$\,keV X-ray luminosity evolution is shown in the bottom panel of Fig.~\ref{pic:lcs} while the count rate variability is plotted in the top panel of Fig.~\ref{pic:countrate_and_ph}. The first flaring episode peaked at a luminosity of $L_{\textsc{X}} \approx 6\times 10^{42}$ \,${\rm erg}\,{\rm s}^{-1}$ in the end of June $2023$ (see Fig.~\ref{pic:lcs}.)  This flare peaked two weeks before the optical peak. The count rate dropped by a factor of $\sim 10$ within a timescale of $\sim 60$\,days. Later, the X-ray emission showed limited variability for several months. A hint of increased emission was detected in mid-February $2024$, which was also followed by a bump in the optical and UV light curves around two weeks later. In the end of April $2024$, a rapid and extreme flaring episode started in X-rays that lasted less than a month but reached a luminosity of $L_{\textsc{X}} \approx 4\times 10^{43}$\,erg/s. which is almost a magnitude larger than the peak of the first X-ray flaring episode. Interestingly, this X-ray flare was also accompanied by a bump in the optical $1-2$ weeks later. A similarly luminous and rapid flare was detected in August $2024$ and once more, accompanied by an optical bump $1-2$ weeks later. Around $90$\,days passed between the second and third and the third and fourth X-ray peaks.

\begin{figure}
  \centering
  {\includegraphics[width=90mm]{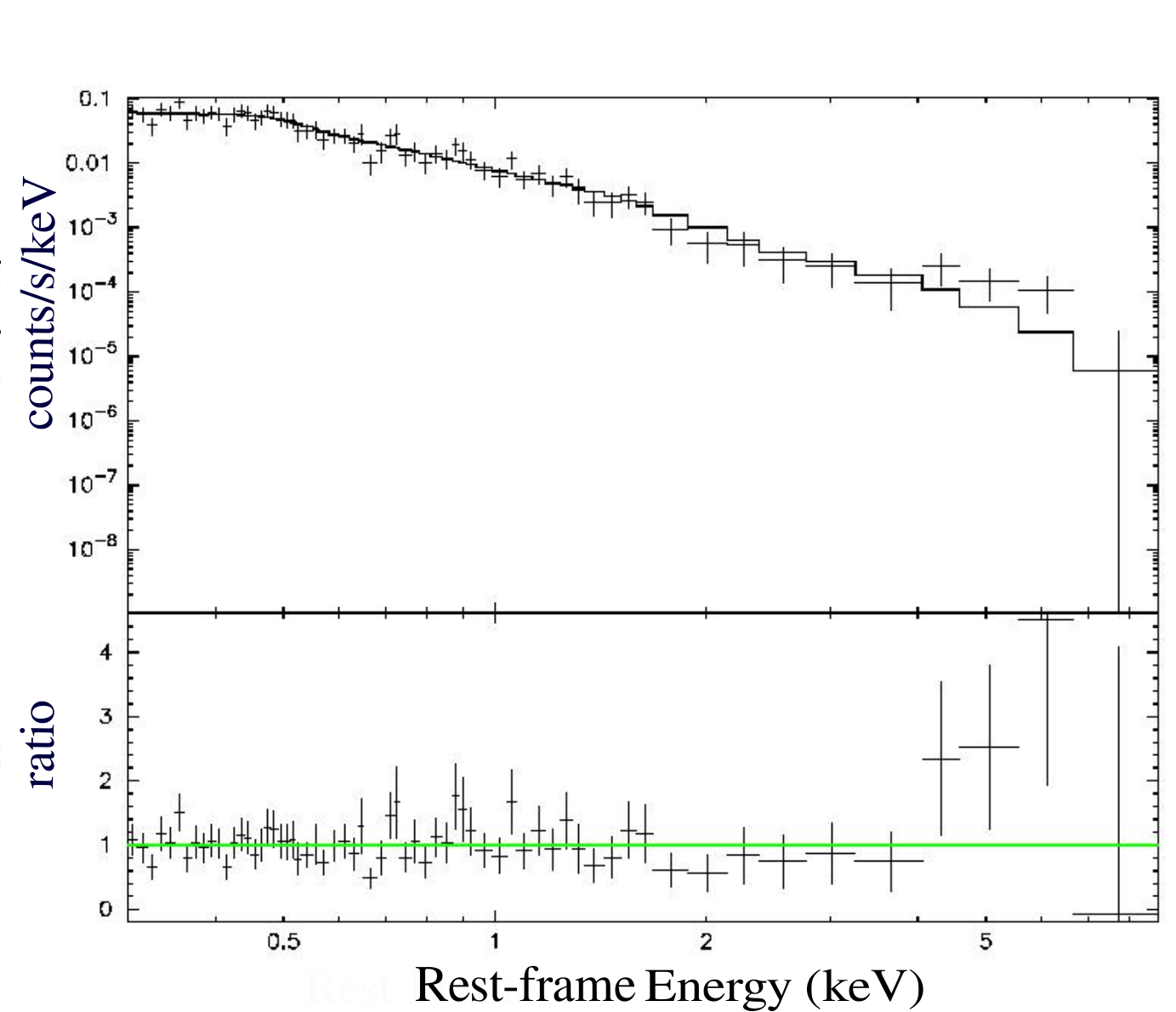}}
  \caption{The overall \textit{Swift}-XRT $0.3-10$\,keV spectrum of AT2019aalc. We binned the spectrum to have a minimum of five counts per spectral bin and fit with an absorbed blackbody$+$power-law model.}
  \label{pic:spectrum_full}
\end{figure}

In Fig.~\ref{pic:spectrum_full} we show the source spectrum derived from the  observing campaign.  The spectral index of the stacked spectrum fitted with a power-law$+$blackbody model is $\Gamma = 3.3 ^{+0.5}_{-0.4}$ while the blackbody temperature is $kT = 94 ^{+6}_{-7}$\,eV. These imply an overall unabsorbed luminosity value of L$_{\textsc{X}} = 2.56^{+0.07}_{-0.07} \times 10^{42}$ erg/s. The mean count rate of the observations is $0.038 \pm 0.001$\,cts s$^{-1}$. The photon index variability of AT2019aalc is presented in the bottom panel of Fig.~\ref{pic:countrate_and_ph}.  

AT2019aalc (= SRGe J152416.7$+$045118) was observed by the eROSITA \citep{2021A&A...656A.132S} soft X-ray telescope four times starting from 2020-02-02 (roughly half a year after the peak of the first optical flare) with a $6$ months cadence. The X-ray light curve reached a plateau between August $2020$ and January $2021$  with a flux of F$_{\textsc{X}}\approx 4.6 \times 10^{-13}$\,erg cm$^{-2}$s$^{-1}$ in the energy range of $0.3-2$\,keV. The source had a soft thermal spectrum described with a blackbody temperature of $kT = 172 \pm 10$\,eV \citep{Lancel}. One observation with XMM-Newton Observatory \citep{2001A&A...365L...1J} was performed on 2021-02-21 during a period of optical quiescence. The best-fit power-law index and the observed $0.3-8$\,keV flux of this observation were $\Gamma = 2.6 ^{+0.1}_{-0.1}$ and F$_{\textsc{X}}\approx 5.1^{+0.15}_{-0.25}\times 10^{-13}$\,erg cm$^{-2}$s$^{-1}$ (68$\%$ confidence), respectively \citep{2023ATel16118....1P}.

\subsection{Radio}
\subsubsection{Temporal evolution}
\label{subsect:temp_radio}
The pre-flare system was detected in the NRAO VLA Sky Survey (NVSS, \citealt{1997ApJ...475..479W}) in February 1995 and in the FIRST survey in February 2000 (both at $1.4$\,GHz) with peak flux densities of $6.0 \pm 0.5$\,mJy and $6.2 \pm 0.1$\,mJy respectively, suggesting no substantial variability.  These flux densities correspond to a monochromatic radio luminosity of $\approx 2 \times 10^{22}$\,W Hz$^{-1}$ measured at $1.4$\,GHz, which is typical for Seyfert galaxies \citep{1984A&A...136..206M}.
 
Later, the transient's sky location was observed in the Karl G. Jansky Very Large Array Sky Survey (VLASS, \citealt{2020PASP..132c5001L}). The VLASS observes the entire sky visible to the VLA at $2-4$\,GHz wide band three times. Our source was detected with peak flux density values of $2.9 \pm 0.2$\,mJy/beam (on 2019-03-14), $5.7 \pm 0.2$\,mJy/beam (on 2021-11-06) and $4.0 \pm 0.2$\,mJy/beam (on 2024-07-20), corresponding to 3~GHz radio luminosities ($\nu L_{\nu}$) of $3\times 10^{38}$\,${\rm erg}\,{\rm s}^{-1}$, $5\times 10^{38}$ \,${\rm erg}\,{\rm s}^{-1}$ and $3.5\times 10^{38}$\,${\rm erg}\,{\rm s}^{-1}$, respectively. This factor of $\approx2$ increase in flux from the first VLASS epoch (three months before the first optical peak) to the second VLASS epoch (2 years post-peak) is statistically significant (at the 8$\sigma$-level, as estimated from the rms in the VLASS ``Quick Look" images, \citealp{Lancel}).

In addition, we found a radio source at the position of AT2019aalc in the first data release of the Rapid ASKAP Continuum Survey \citep[RACS:][]{racs1}. The source was observed in two different fields and, consequently, at two different epochs. The observed peak flux densities of $9.2 \pm 0.5$\,mJy/beam (on 2019-04-24) and $9.5 \pm 0.5$\,mJy/beam (on 2020-04-30) correspond to a $888$\,MHz radio luminosity of $\approx 2.5\times 10^{38} \,{\rm erg}\,{\rm s}^{-1}$ and indicate that the radio flare observed in the VLASS data started at least $300$\,days after the peak of the first optical flare, sometime between April $2020$ and November $2021$.

The radio luminosity in the beginning of our ATCA monitoring was at the same level ($5\times 10^{38}$ \,${\rm erg}\,{\rm s}^{-1}$) when the source was observed for the second time in the VLASS in April 2021. Since the radio brightness does not show rapid variability during the months of our ATCA monitoring, we can reasonably assume that the sharp increase resulted in a long-lasting plateau. We present the ATCA radio multi-frequency light curves of AT2019aalc together with the archival observations in the middle panel of Fig.~\ref{pic:lcs}.

\subsubsection{The radio spectrum}

To further characterize the radio emission of AT2019aalc, we derived the radio spectral index of the radio source in each epoch of our ATCA monitoring. The spectral index between $2.1$\,GHz and $9.0$\,GHz ($\alpha$, defined as $S \propto \nu^{\alpha}$) does not show any obvious signs of variability on a timescale of $9$\,months. We give an average spectral index of $\alpha^{\textrm{9GHz}}_{\textrm{2.1GHz}} \approx -1.1$. We calculated a pre-flare spectral index using the VLASS epoch $1$ and RACS epoch $1$ flux densities. The spectral index we derived this way is $\alpha^{\textrm{3GHz}}_{\textrm{888MHz}} = -0.95 \pm 0.07$.

Our high-frequency K-band observations centered at $18$\,GHz, however, do not follow the steep power-law and suggest a turnover between $9$\,GHz and $18$\,GHz. 

\begin{figure}
  \centering
  {\includegraphics[width=80mm]{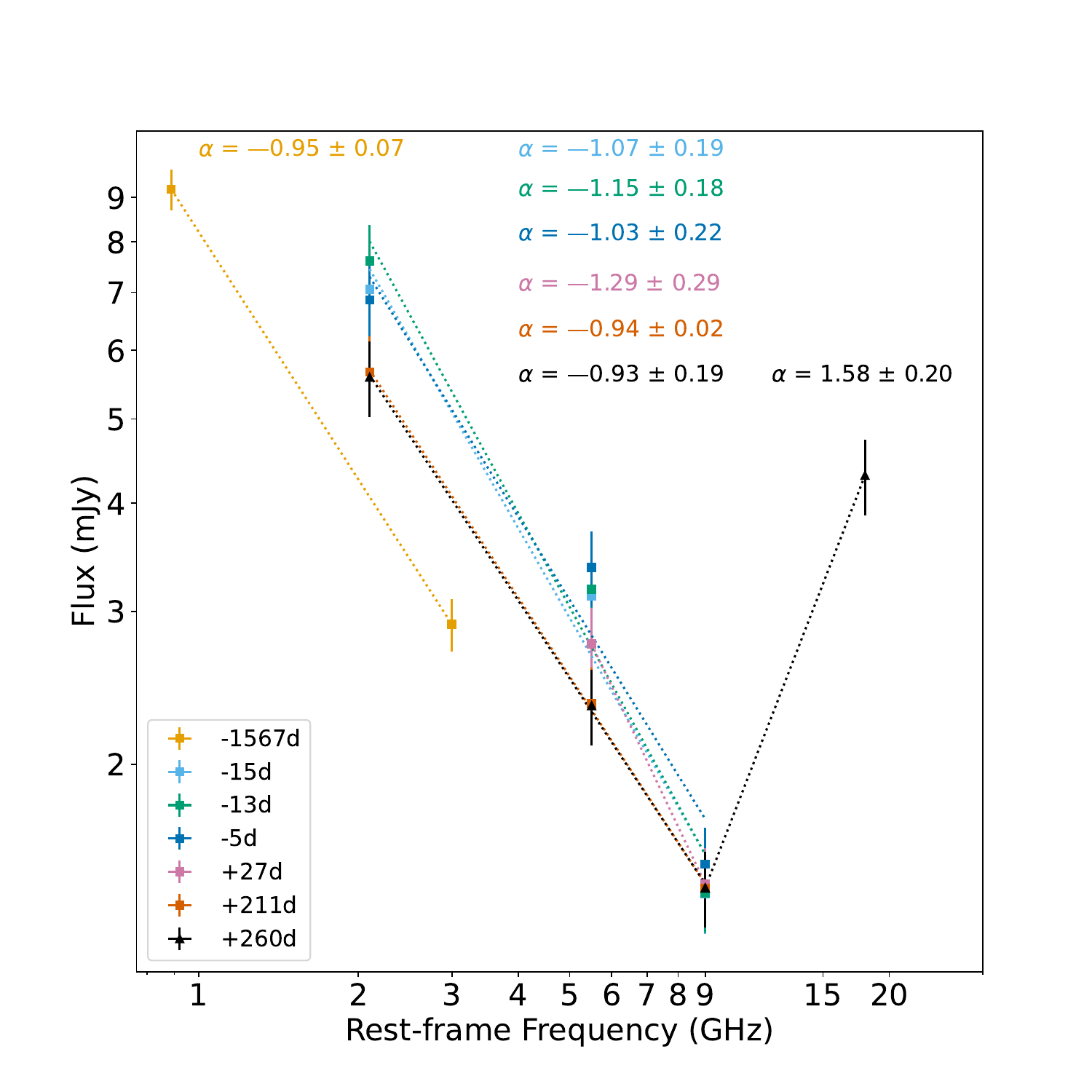}}
  \caption{Radio spectra of AT2019aalc spanning $3$–$18$\,GHz obtained through our ATCA monitoring. We derived a prior radio flare spectrum between $888$\,MHz and $3$\,GHz from archival fluxes. The times in the legend are relative to the peak of the second optical flare.}
  \label{pic:atca_sp}
\end{figure}

\subsubsection{VLBI detection}

\begin{figure}
  \centering
  {\includegraphics[width=90mm]{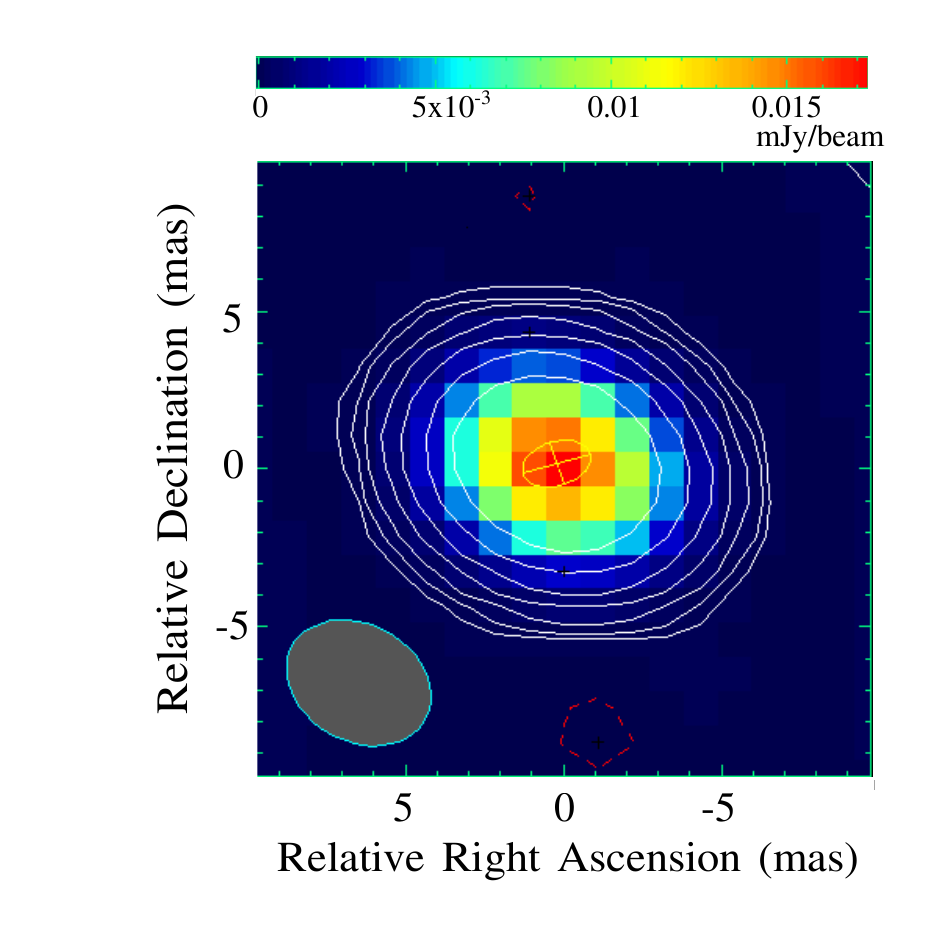}}
  \caption{Naturally weighted $1.7$ GHz EVN+e-MERLIN high-resolution VLBI map of AT2019aalc. The peak brightness is $17.3\textrm{\,mJy\,beam}^{-1}$. The grey ellipse in the lower-left corner represents the Gaussian restoring beam. Its parameters are $4.82\textrm{\,mas} \times 3.7$\,mas (FWHM) at a major axis position angle of PA $=60\fdg8$. The lowest contour level is drawn at $\pm 3\sigma$ image noise level corresponding to $0.08\textrm{\,mJy\,beam}^{-1}$. Further positive contour levels increase by a factor of $2$. The red dashed contours represent the negative contours. The yellow ellipse is the fitted Gaussian modelfit component used to describe the brightness distribution of the radio source.}
  \label{pic:evn}
\end{figure}

The naturally-weighted VLBI map of AT2019aalc is presented in Fig.~\ref{pic:evn}. We detected a single radio-emitting feature with a signal-to-noise ratio (SNR) of $\approx 19 \sigma$. We obtained the following parameters of the fitted component: a total flux density of $S=21 \pm 1.8$\,mJy and a size of $d = 2.23\times1.39 \pm 0.3 \times 0.3$\,mas at a position angle of $\phi = -73.3\degr$. We calculated the uncertainties of the flux density and size parameters following the formulas of \cite{fomalont}. The minimum resolvable size ($\theta_{\textrm{lim}}$) of a Gaussian component fitted to naturally weighted VLBI data was calculated following \cite{2005AJ....130.2473K}. This yields $\theta_{\textrm{lim}} = 1.05 \times 0.81$\,mas, implying that the fitted Gaussian component is resolved.

We calculated the brightness temperature of the radio-emitting feature found at the position of AT2019aalc using the following equation \citep{Tb, Tb2}: 
\begin{equation}
T_{\mathrm{b}} = 1.22 \times 10^{12} (1+z) \frac{S}{\theta_{\mathrm{maj}} \times \theta_{\mathrm{min}} \nu^{2}}\,\textrm{K,}
\end{equation}
where $S$ is the flux density of the fitted Gaussian component measured in Jy, $\theta_{\mathrm{maj}}$ and $\theta_{\mathrm{min}}$ are the major and minor axes full width at half maximum (FWHM) of the fitted component in mas, and $\nu$ is the observing frequency in GHz. The obtained brightness temperature is 
$T_{\mathrm{b}} = (3.0 \pm 0.8) \times 10^{8}$\,K at $1.7$\,GHz. The brightness temperature is an effective parameter that is commonly used in radio astronomy to describe the physical properties of emitting material in astrophysical objects \citep[e.g.,][]{2015A&A...574A..84L}. 

\subsection{Spectroscopic results}
\label{subsect:spectro_res}
We present the archival and the newly obtained optical spectra of AT2019aalc in Fig.~\ref{fig:a1}, marking the notable emission lines. In the following, we highlight the remarkable line detection of the pre-flare spectrum observed before the first optical flare, the one observed between the two flares (post-flare-1 spectrum) and the five spectra taken during the second flare (over-flare-2 spectra).

\subsubsection{Pre-flare spectrum}

A spectrum of the pre-flare system was observed in April $2008$ by the SDSS spectrograph. We present this spectrum with the best fits for the stellar and gas components in Fig.~\ref{fig:a2}. 
\begin{itemize}
    \item We identify strong Balmer emission lines. We calculated FWHM(Balmer)$\approx2800$\,kms$^{-1}$ for the broad component of the Balmer lines in the pre-flare spectrum, which is typical for an unobscured, broad-line AGN. The line widths (FWHM(Balmer)$\approx2600$\,km s$^{-1}$) presented in the SDSS DR7 broad-line AGN catalog \citep{2019ApJS..243...21L} for our source are consistent with this classification.
    \item Several forbidden lines such as [O\,{\sc iii}] $\lambda 4959$ and [O\,{\sc iii}] $\lambda 5007$, [S\,{\sc ii}] $\lambda 6716$, [S\,{\sc ii}] $\lambda 6731$ and [N\,{\sc ii}] $\lambda 6584$ were detected. These lines are produced in the narrow-line region (NLR) in AGN. We detected the [O\,{\sc ii}] $\lambda 3726 +$ [O\,{\sc ii}] $\lambda 3729$ line doublet, which is a potential indicator of AGN-driven outflows \citep[e.g.,][]{2020A&A...644A..54S}.
    \item The pre-flare spectrum also indicates the detections of two high-ionization coronal lines ([Fe\,{\sc x}]$\lambda 6375$ and [Fe\,{\sc xiv}]$\lambda 5303$) which are further discussed below.
\end{itemize}

\subsubsection{Post-flare-1 spectrum}
\label{subsect:post_flare_sp}

The first LRIS spectrum was observed almost two years after the first optical flare and two years before the second one. 
\begin{itemize}
    \item In this spectrum, prominent Balmer lines were detected, similarly to the SDSS spectrum. The spectrum exhibits the aforementioned forbidden lines detected in the archival spectrum.
    \item  We identify the Bowen Fluorescence lines O\,{\sc iii} $\lambda 3133$ and He\,{\sc ii} $\lambda3203$. 
    \item The high ionization coronal lines [Fe\,{\sc vii}]$\lambda 6087$, [Fe\,{\sc x}]$\lambda 6375$, [Fe\,{\sc xiv}]$\lambda 5303$ and [Fe\,{\sc vii]$\lambda 5721$} were detected. Furthermore, we identify the [Ne\,{\sc V}] emission lines at $3345 \AA$ and $3426 \AA$ in the spectrum. Note that the pre-flare SDSS spectrum does not cover these wavelengths, therefore these lines are not necessarily related to the optical flares discussed here. These neon lines are prominent emission lines originating from the inner NLR excited by the AGN via photoionization or shocks, and are sometimes considered as coronal lines \citep[e.g.,][]{2020ApJ...895..147C}.
\end{itemize}

\subsubsection{Over-flare-2 spectra}
\begin{figure*}
  \centering
  {\includegraphics[width=\textwidth]{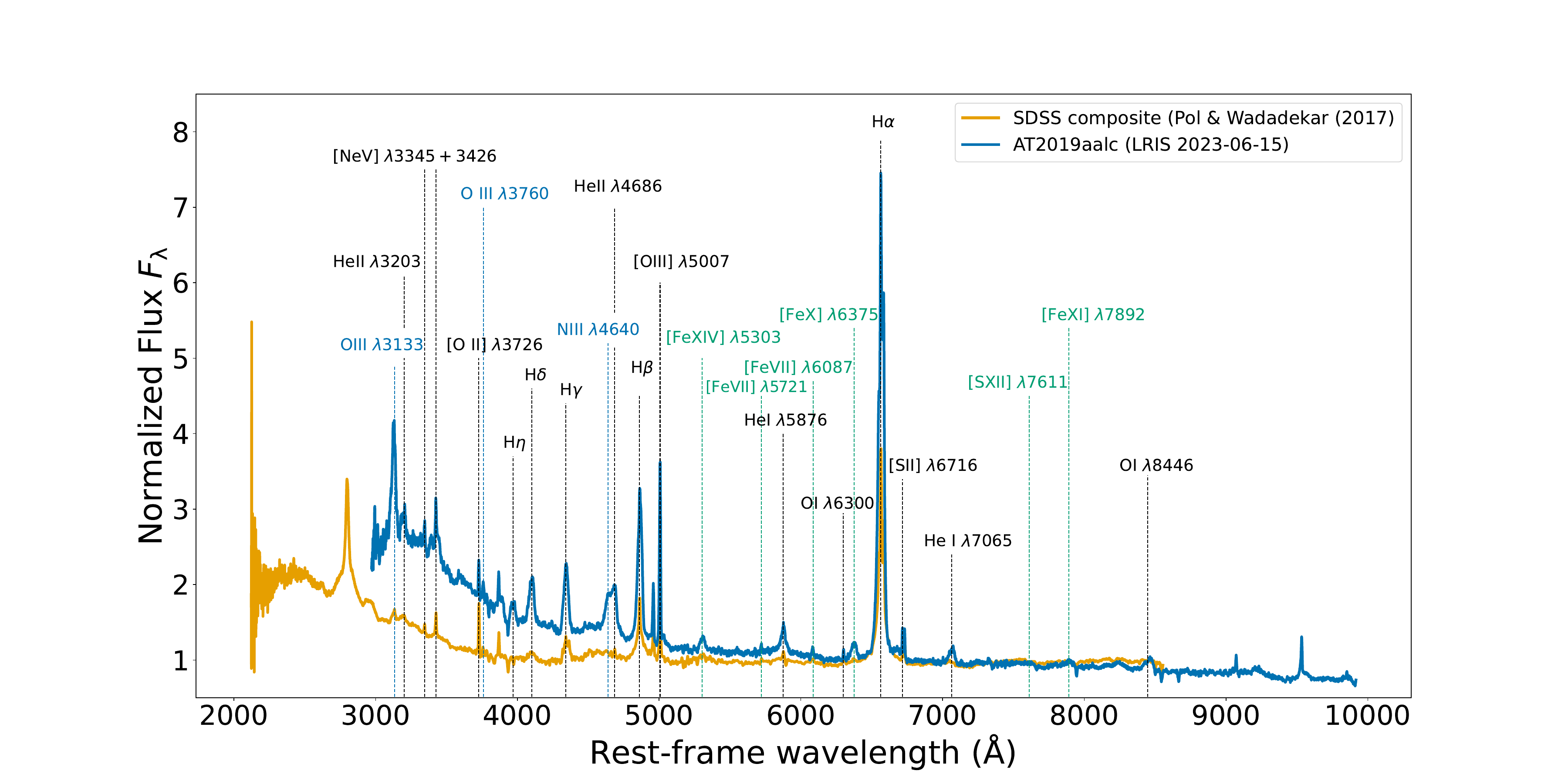}}
  \caption{The over-flare-2 LRIS spectrum of AT2019aalc in comparison with the SDSS composite spectrum of $10112$ Seyfert 1 galaxies published in \cite{2017MNRAS.465...95P} and normalized between $7400$ and $7500 \AA$. The most remarkable lines are shown with vertical lines. The most notable differences are the appearance of the Bowen lines (marked with blue) and the high-ionization coronal lines (marked with green) in the spectrum of AT2019aalc.}
  \label{pic:spectra_comp}
\end{figure*}

We performed five optical spectroscopic observations of AT2019aalc during the second flare. These are characterized by a blue continuum with several more emission lines than the spectra discussed above. We compare the second LRIS spectrum of AT2019aalc (covering the broadest wavelength range) with a composite SDSS spectrum based on $10112$ Seyfert 1 galaxies published in \cite{2017MNRAS.465...95P}, in Fig.~\ref{pic:spectra_comp}. This comparison shows similarities but also remarkable differences discussed below.
\begin{itemize}
    \item  Over the second optical flare, strong Balmer lines were detected, with two of them (H$\delta$ and H$\eta$) appearing only in the over-flare-2 spectra. 
    \item Over the second optical flare of AT2019aalc, we identified the Bowen fluorescence line complex N\,{\sc iii} $\lambda 4640$+He\,{\sc ii} $\lambda 4686$. The complex was not detected in the earlier spectra. The relative strength of the BF complex to H$\beta$ reached a maximum on 2023-07-11, very close to the optical continuum peak (2023-07-13). As we detected the Bowen line N\,{\sc iii} $\lambda 4640$ even $315$\,days after the optical peak of the second flare, we conclude that it persists over long timescales. We also find hints of O\,{\sc iii}$ \lambda 3760$, which is also known as a Bowen line \citep{2007A&A...464..715S}, however, due to a possible blending from Fe\,{\sc vii}$ \lambda 3759$, its detection is questionable. Nevertheless, this Bowen line is created through the same channel (O1) as the above-discussed O\,{\sc iii} $\lambda 3133$, which supports the hint of detection. O\,{\sc iii}$ \lambda 3760$ was detected in some metal-rich TDEs before \citep{2019ApJ...887..218L}, such as AT2019dsg \citep{2021MNRAS.504..792C}. 
    \item We detect the high-ionization coronal lines idenfitied in the post-flare-1 LRIS spectrum over the second flare in each spectrum. Furthermore, a new coronal line appears: [Fe\,{\sc xi}] $\lambda7892$, which has an ionization potential of $262\,\mathrm{eV}$ \citep[e.g.,][]{2022ApJ...936..140R}. We also detected the [Ne\,{\sc V}] emission lines at $3345 \AA$ and $3426 \AA$ in the second LRIS spectrum, as well as in the post-flare-1 spectrum.
\end{itemize}

\section{Discussion}
\label{sect_5}

\subsection{Spectroscopic properties}
\label{subsect:spectro}

\subsubsection{Balmer lines}
AT2019aalc exhibits prominent Balmer lines in each spectrum. In addition, a more complex structure of these lines appears in the late-time over-flare-2 spectrum. The Balmer series lines here exhibit a redward wing which however, in some cases, might result from blending with other lines, such as H$\delta$ and the BF line N {\sc iii $\lambda 4103$}. Interestingly, the post-flare-I spectrum shows a similar structure to the Balmer line series.

\subsubsection{Bowen fluorescence lines}
Wavelength coincidences between emission lines ("line fluorescence") can be an important source of radiative excitation. The Bowen Fluorescence mechanism in astrophysical environments is a good indicator of absorbed EUV$-$soft X-ray flux (at wavelengths shorter than the He\,{\sc ii} Lyman limit of $h\nu \geq 54.4$\,eV), as these high-energy photons are converted into the He\,{\sc ii} Lyman$\alpha \lambda 303.782 \AA$ emission required for the excitation of the O\,{\sc iii} and N\,{\sc iii} lines in the optical and NUV regimes  \citep{1985ApJ...299..752N,1990agn..conf...57N}. This is consistent with the multiple soft X-ray flares and the extreme UV luminosities of AT2019aalc. The BF line O\,{\sc iii} $\lambda 3133$ has been observed in the Bowen Fluorescence Flares AT2017bgt \citep{Trakhtenbrot} and AT2021loi and only in a few Seyfert galaxies before \citep[e.g.,][]{1986ApJ...310..679M,1990ApJ...362...74S}. However, we note that due to the atmospheric cutoff at $\approx 3100 \AA$ these lines are difficult to detect. This line is typically weaker in regular Seyfert galaxies than in the case of AT2019aalc (see Fig. \ref{pic:spectra_comp}). The BF line N\,{\sc iii} $\lambda 4640$ is not generally seen in AGN \citep[e.g.,][]{2001AJ....122..549V}, however, it has been observed in several optical TDEs \citep{spectra2} and BFFs \citep{Trakhtenbrot,AT2021loi}.  

\begin{figure}
  \centering
  {\includegraphics[width=80mm]{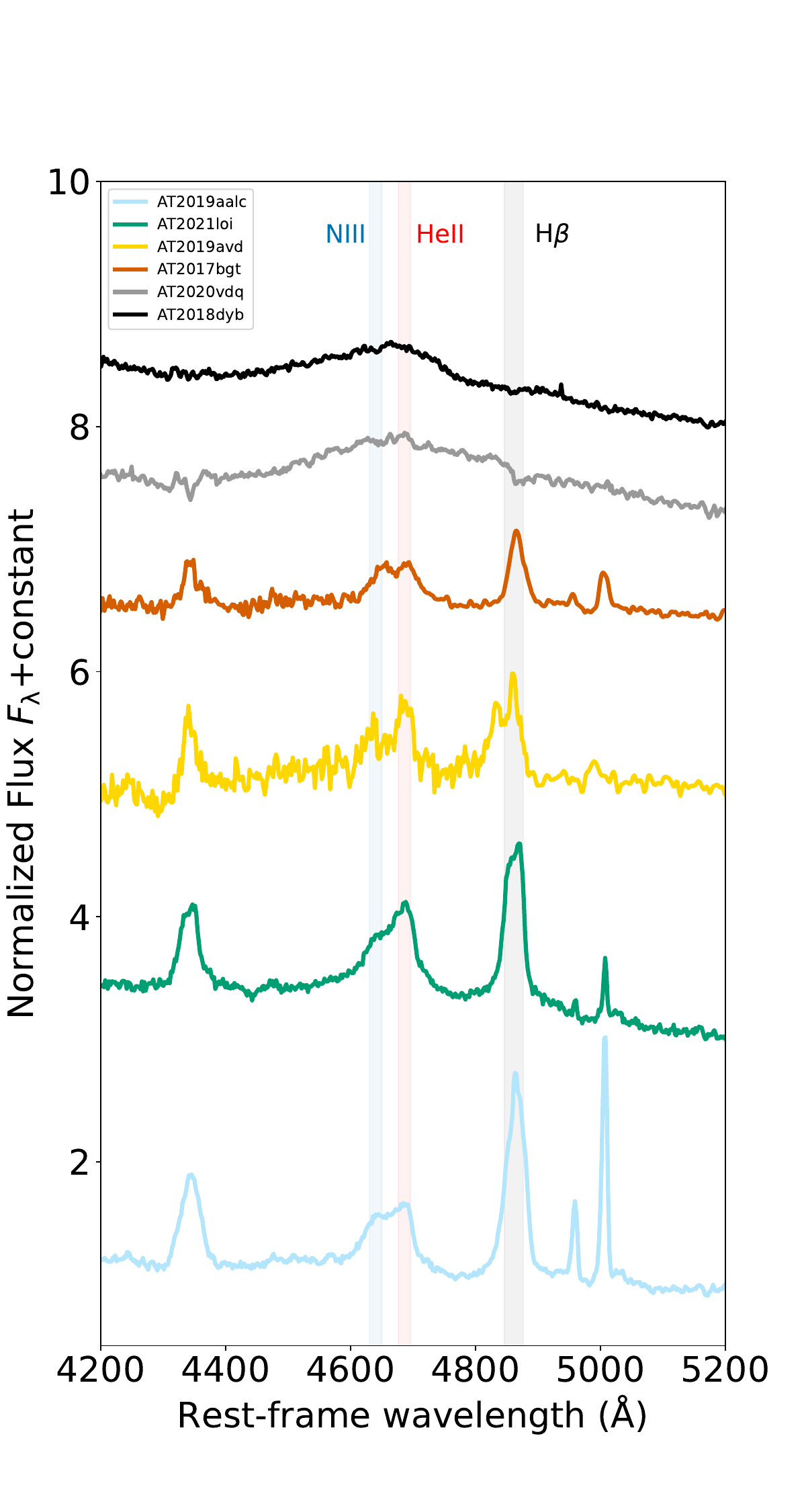}}
  \caption{The second LRIS spectrum of AT2019aalc zoomed in on the He\,{\sc ii} $\lambda 4686+$N\,{\sc iii} $\lambda 4640$ complex region, compared to those of the BFFs AT2021loi \citep{AT2021loi} and AT2017bgt \citep{Trakhtenbrot}, the peculiar transient event AT2019avd \citep{2020TNSAN.105....1T}, the repeating TDE AT2020vdq \citep{2020vdq} and the TDE-Bowen AT2018dyb \citep{2018TNSCR.998....1P}. All of the plotted spectra were normalized at $5200 \AA$.}
  \label{pic:spectra_comp_classes}
\end{figure}

\begin{figure}
  \centering
  {\includegraphics[width=90mm]{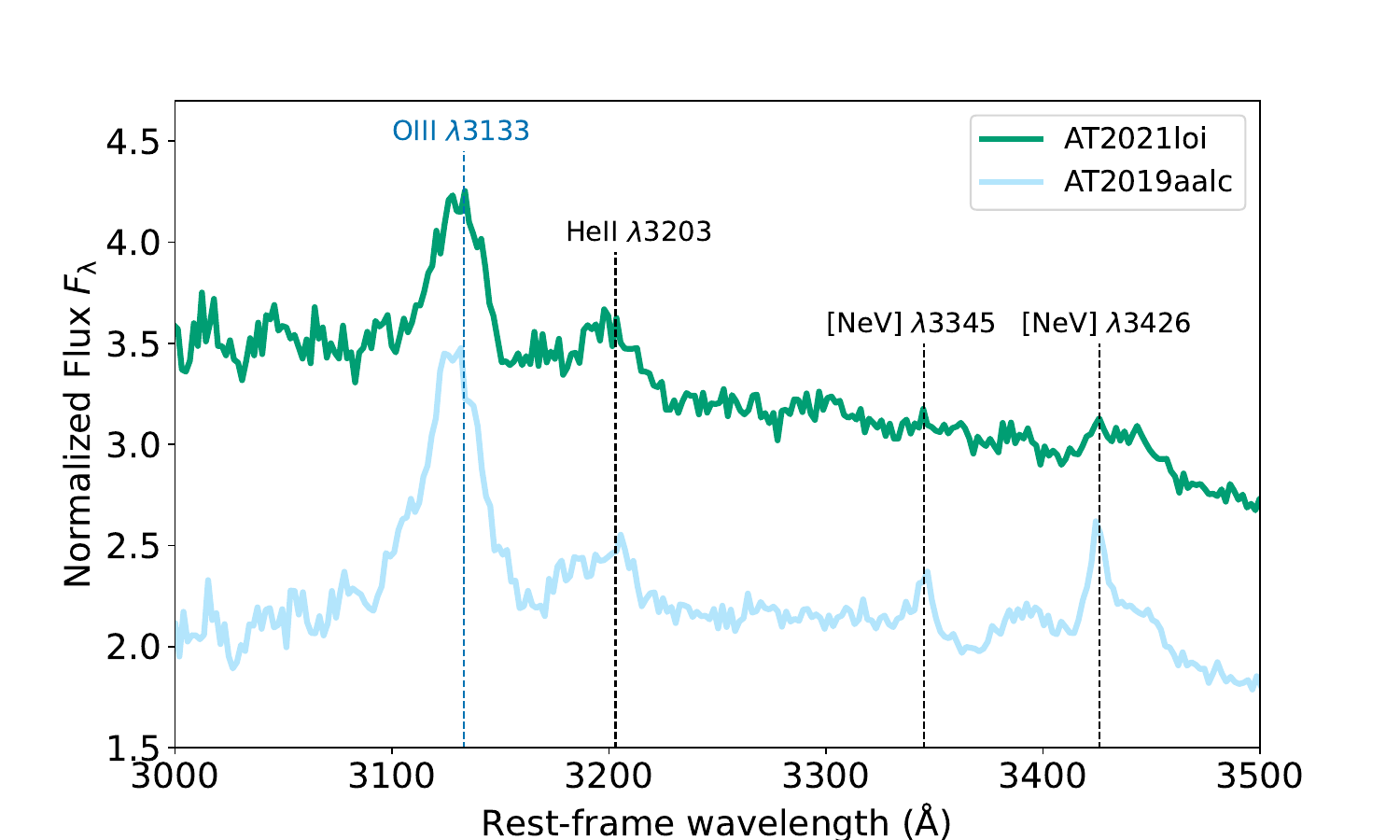}}
  \caption{The second LRIS spectrum of AT2019aalc zoomed in on the O\,{\sc iii} $\lambda 3133+$He\,{\sc ii} $\lambda 3203$ complex region, compared to that of the BFF AT2021loi published by \cite{AT2021loi}. The spectra were normalized at $5200 \AA$.}
  \label{pic:spectra_comp_classes_loi}
\end{figure}

The detection of a BF line almost two years after the first optical flare of AT2019aalc suggests its long persistence, similar to the BFFs studied by \cite{Trakhtenbrot}. In the case of AT2017bgt, this line was still detected $470$\,days after the discovery of the transient. The Bowen features in F01004-2237 were detected even $\approx8$\,years after the initial flare \citep{f01}.  We compare the over-flare-2 high-resolution LRIS spectrum of AT2019aalc to the spectra of BFFs, a TDE-Bowen, and a repeating partial TDE in Fig.~\ref{pic:spectra_comp_classes} around the N\,{\sc iii} $\lambda 4640+$He\,{\sc ii} $\lambda 4686$ complex. We also compare the second LRIS spectrum of AT2019aalc with the LRIS spectrum of AT2021loi in Fig.~\ref{pic:spectra_comp_classes_loi}, zooming into the O\,{\sc iii} $\lambda 3133+$He\,{\sc ii} $\lambda 3203$ region.

The He\,{\sc ii $\lambda 4686$} emission line is commonly detected in AGN and known as a recombination line, tracing ionized gas in the vicinity of the central SMBH responsible for the photoionization \citep[e.g.][]{2020MNRAS.498.4550W}. This line is known to be created by photoionization due to (soft) X-ray photons with a rough correspondence of $1$ He\,{\sc ii} photon emitted for each $0.3-10$\,keV X-ray photon \citep{1986Natur.322..511P,2019A&A...622L..10S,2021MNRAS.504..792C}. In the spectra of AT2019aalc, He\,{\sc ii} $\lambda 4686$ is detected together with strong N\,{\sc iii} and O\,{\sc iii} BF lines and  He\,{\sc ii} $\lambda 3203$, indicating its relation to the Bowen Fluorescence mechanism. Furthermore, we calculated a maximum line intensity ratio of $F(\textrm{He\,{\sc ii}})/F(\textrm{H}\beta) \approx 0.7$, which significantly exceeds the typical ratios estimated for AGN (earlier \cite{2001AJ....122..549V} estimated ratios of $F(\textrm{He\,{\sc ii}})/F(\textrm{H}\beta) \leq 0.05$ while \cite{2012MNRAS.421.1043S} gives $F(\textrm{He\,{\sc ii}})/F(\textrm{H}\beta) \leq 0.1$). Our value is, however, more comparable with the line ratio estimated for the BFF AT2017bgt ($F(\textrm{He\,{\sc ii}})/F(\textrm{H}\beta) \approx 0.5$, \citealp{Trakhtenbrot}).

\subsubsection{Coronal lines}
\label{subsect:coronal}
 
Based on a study of SDSS spectra, \cite{2012ApJ...749..115W} reported the discovery of seven Extreme Coronal Line Emitter (ECLE) galaxies, which show extremely strong coronal lines. Numerous TDEs were also found to exhibit coronal lines. The BFF AT2021loi \citep{AT2021loi} and the spectacular transient event AT2019avd \citep{avd1} also exhibit strong coronal lines. The BFF F01004-2237 showed variable coronal lines on a timescale of $3$\,years \citep{f01}. Based on the publicly available spectra of AT2017bgt on the Transient Name Server\footnote{\url{https://www.wis-tns.org/}} (TNS) we detect enhanced coronal line emission in this BFF, too. 

The most plausible origin of the ECLEs is that they represent echoes of a past TDE outburst \citep{2012ApJ...749..115W,2024MNRAS.528.4775H,2024MNRAS.532..112K,2024ApJ...977..258N}. Notably, \cite{2023MNRAS.525.1568S} revealed that the TDE AT2019qiz more likely follows the line ratio correlations of ECLEs instead of regular AGN. The coronal lines in AT2019qiz appeared around $400$\,days after the optical peak, further supporting this scenario. Moreover, the estimated rate of coronal line emitters is consistent with the estimated rate of TDEs \citep{2012ApJ...749..115W}. Later, \cite{ecle_rate_callow} discovered 9 new ECLEs using SDSS spectra and refined the previous rate calculations. They found that the ECLE rates are below the TDE rates, however, can still be consistent with the assumption that only $10$ to $40$ percent of all TDEs produce variable coronal lines. According to \cite{2012ApJ...749..115W}, the origin of the coronal lines in ECLEs can be explained with the liberation of Fe from dust grains that were destroyed by the echo of past TDE flares. A follow-up spectroscopic analysis of the original ECLE sample of \cite{2012ApJ...749..115W} revealed that the coronal line signatures did not recur in five out of the seven objects, supporting their transient light echo origin \citep{ecle_mw}. The large dust echo of AT2019aalc is consistent with this picture. Furthermore, the TDEs with coronal line detection have more luminous IR dust echoes compared to other optical TDEs. AT2019qiz is one of the few TDEs included in the accretion flare sample of \cite{Lancel}, due to its strong dust echo. Another coronal line-detected TDE AT2017gge is part of the mid-infrared (MIR) dust echo flare sample of \cite{2024MNRAS.531.2603H}, which contains $19$ ANTs with high dust covering factors. The extreme dust echo seems to characterize the BFFs as well. The BFFs AT2021loi, AT2017bgt, OGLE17aaj and the candidate BFF AT2019avd are present in the above-mentioned MIR dust echo sample \citep{2024MNRAS.531.2603H} and two of them are also part of the Flaires sample of dust-echo-like IR flares \citep{2025A&A...695A.228N}. The BFF F01004-2237 exhibited a MIR flare \citep{2017ApJ...841L...8D}. The Bowen line transient AT2019pev shows strong high-ionization coronal lines \citep{2021ApJ...920...56F,2022MNRAS.515.5198Y} and is also part of the MIR flare sample of \cite{2024MNRAS.531.2603H}. These findings suggest a connection not only between the ECLEs and TDEs, but also the BFFs. 

\begin{figure}
  \centering
  {\includegraphics[width=90mm]{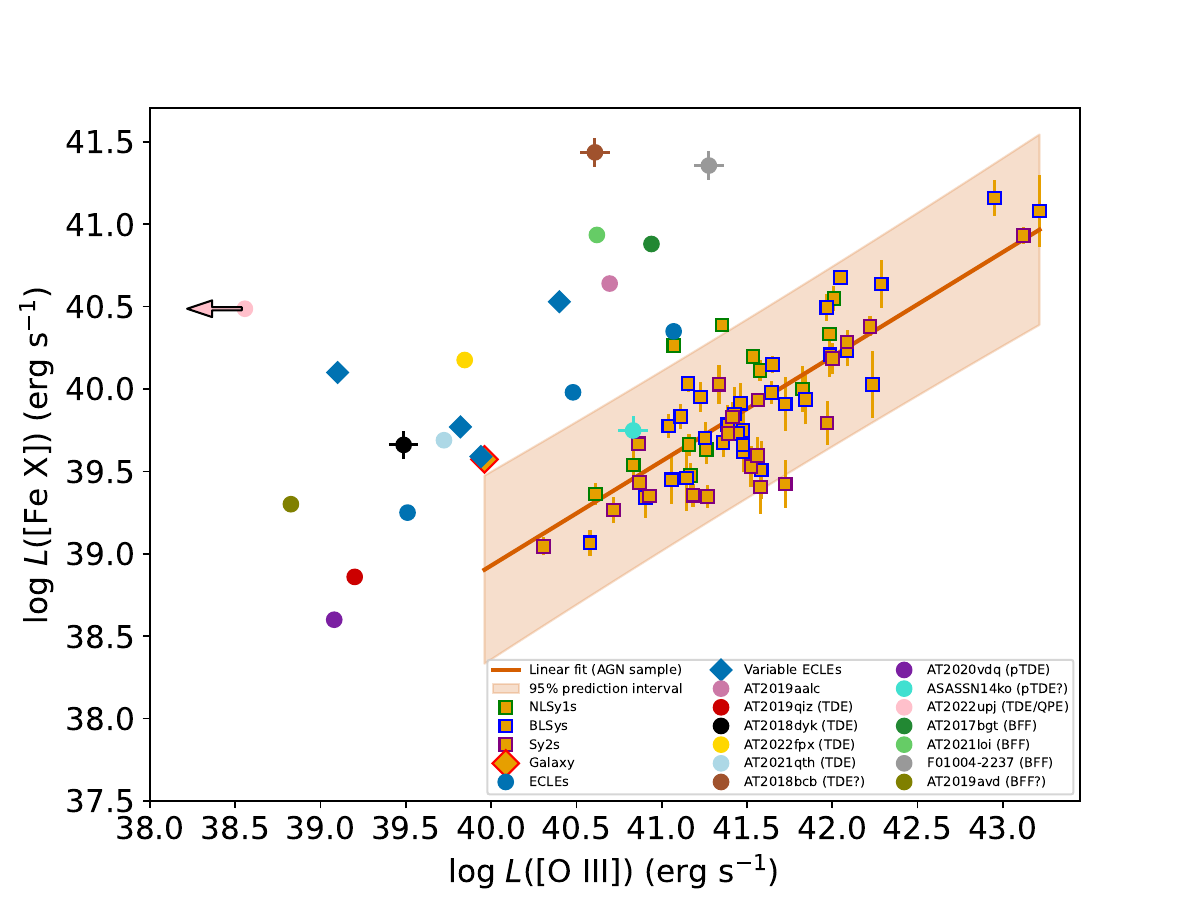}}
  \caption{[Fe\,{\sc x}]$ \lambda 6375$ versus [O\,{\sc iii}$] \lambda 5007$ luminosities of ECLEs \citep{2012ApJ...749..115W}, variable ECLEs \citep{2013ApJ...774...46Y} AT2019aalc (this work), the TDEs AT2019qiz \citep{2023MNRAS.525.1568S}, AT2018dyk, AT2022fpx \citep{2025ApJ...990...22L}, AT2018bcb, AT2021qth and the TDE/QPE AT2022upj, the repeating pTDEs AT2020vdq \citep{2025ApJ...983..159S} and ASASSN-14ko, the BFFs AT2017bgt, AT2021loi \citep{AT2021loi} and F01004-2237 and the BFF candidate AT2019avd \citep{avd1}. A sample of different types of Seyfert galaxies \citep{2009MNRAS.397..172G} is plotted as well. The orange line represents the a linear fit of the sample and the orange shaded area represents the 95\% prediction interval of it. The transient sources have clearly higher [Fe\,{\sc x}]/[O\,{\sc iii}] luminosity ratios than the Seyfert sample.}
  \label{pic:coronal}
\end{figure}

Fig.~\ref{pic:coronal} shows the [Fe\,{\sc x}]$ \lambda 6375$ versus the [O\,{\sc iii}$] \lambda 5007$ luminosities of the ECLEs, the coronal-line detected TDEs and BFFs and AT2019aalc, together with the coronal-line detected SDSS Seyfert sample of \cite{2009MNRAS.397..172G}. For this plot, we performed modeling of the publicly available spectra\footnote{We retrieved the LRIS spectra from the Transient Name Server (TNS; \url{https://www.wis-tns.org}) and the ESO Faint Object Spectrograph and Camera version 2 (EFOSC2) spectra from the ESO Science Portal (\url{archive.eso.org/scienceportal})} of several TDEs and BFFs using \textsc{PyQSOFit}, following the procedure described in Subsect. \ref{subsect:spectro_data}. These are the TDEs: AT2017gge (2017-09-14, EFOSC2); AT2018dyk (2018-08-08, LRIS); AT2018bcb (2018-05-06, EFOSC2); AT2021qth (2022-05-26, LRIS); AT2022upj (2022-10-26, DBSP); and the BFFs: AT2017bgt (2017-02-23, LRIS) and F01004-2237 (2018-08-13, EFOSC2). Absolute fluxes, and hence derived luminosities, are subject to systematic uncertainties from slit losses and atmospheric extinction differences. We adopt a conservative $\pm20$\% uncertainty on the flux scale to account for these effects, which is combined in quadrature with the statistical uncertainties from the \textsc{PyQSOFit} spectral fits. This method is applied only to cases where the slit position angle differed at least five degrees from the parallactic angle. Fig.~\ref{pic:coronal} clearly implies an offset of the ECLEs, BFFs and AT2019aalc from the regular AGN sample. We note that the only object in the sample of \cite{2009MNRAS.397..172G} classified as a galaxy rather than an AGN is, in fact, the same source as one of the variable ECLEs (which appear as nearly overlapping points in our plot). No publicly available spectra were found of the TDEs AT2021dms, AT2024mvz and AT2021acak while in the case of AT2017gge, it was not possible to safely distinguish [Fe\,{\sc x}]$ \lambda 6375$ from H$\alpha$. We note that AT2021acak exhibited two distinct optical flares \citep{2023RAA....23b5012L}, with the second being more luminous, and showed a He\,{\sc ii} $\lambda 4686+$N\,{\sc iii} $\lambda 4640$ complex similar to that of AT2019aalc.

\subsection{Optical re-brightening}
The long-term ZTF light curve of AT2019aalc shows two well-separated flares. Both flares rose quickly and reached their peaks within two months, followed by a slow decay on yearly timescales. Generally, TDEs are characterized by smoothly and rapidly declining light curves ($F \propto t^{-5/3}$) as expected from mass fallback considerations \citep{1988Natur.333..523R}, and without any prominent bumps in the case of full disruption. AT2019aalc in turn has a slowly decaying first flare with a power-law fit returning a power-law index of $b \approx -0.45$ and a second flare that began four years after the initial flare. The first systematically identified repeating partial TDE (AT2020vdq), however, shows not just a double-peaked long-term optical light curve, but its second flare is $\approx 3$\,times more luminous compared to the first one \citep{2020vdq}. This implies an even more significant optical re-brightening than the one detected in AT2019aalc. In contrast, the candidate pTDE AT2022dbl exhibits with a brighter first flare. The second flare of partial TDEs is expected to follow a steep declining with a power-law of $t^{-9/4}$. In the case of AT2020vdq and AT2022dbl, the second flares declined slower than this \citep{2020vdq,2025ApJ...987L..20M}, but still faster than the second flare of AT2019aalc, which declined with a power-law index of $b \approx -0.23$. Notably, AT2020vdq and AT2022dbl both exhibit slower-declining second flares compared to their first flares, similar to AT2019aalc.

\begin{figure}
  {\includegraphics[width=93mm]{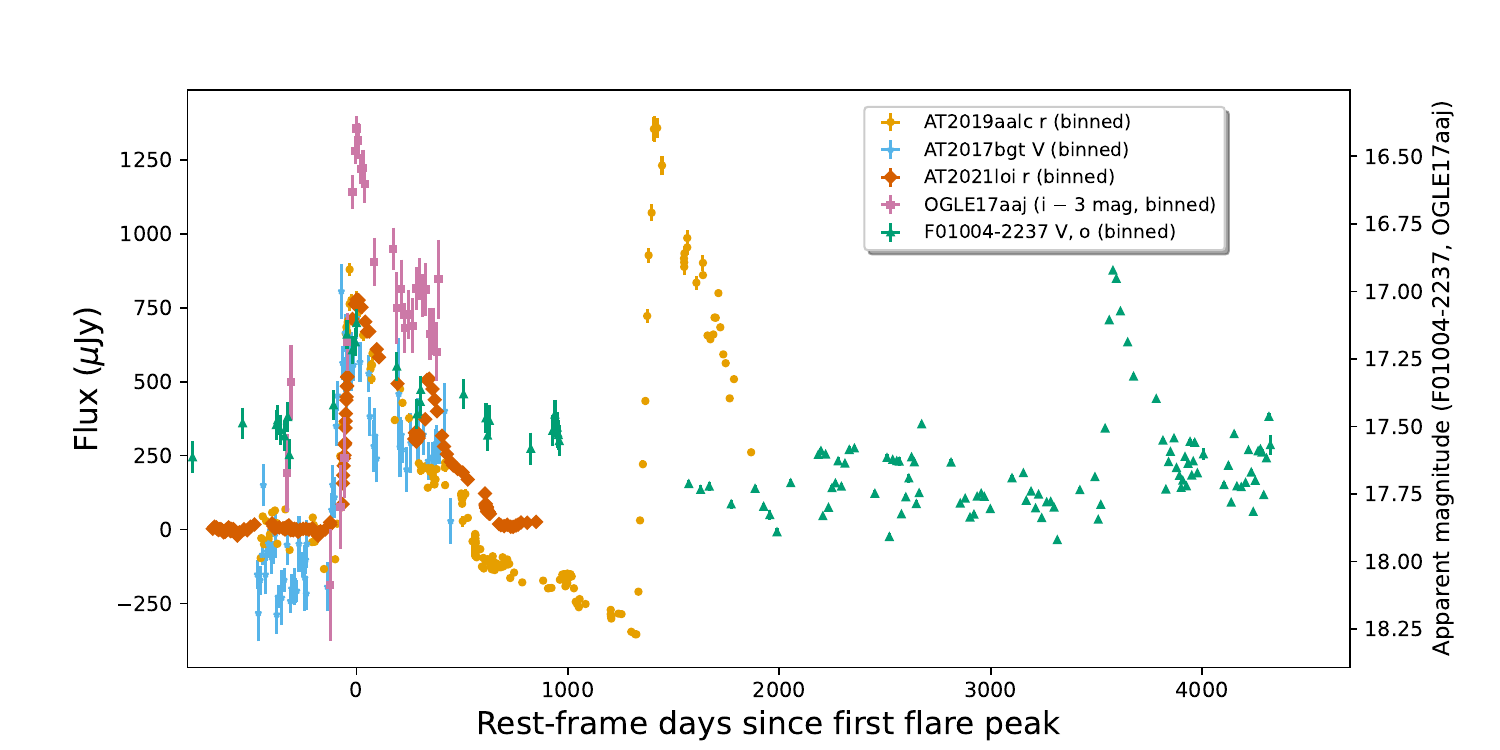}}
  \caption{Long-term optical light curves of AT2019aalc and three classified BFFs. The light curves indicate that optical re-brightening and bumps may be a common property of BFFs: yet, the various sources differ in when these episodes start with regard to the initial flares.}
  \label{pic:comp_lcs}
\end{figure}

Interestingly, most of the sources of the small sample of classified BFFs show optical re-brightening or bumps. Fig.~\ref{pic:comp_lcs} shows the optical evolution of the three BFFs with detected re-brightening episodes, together with AT2019aalc. The optical light curve of AT2017bgt shows a bump roughly $400$\,days after its initial flare which is also clearly visible in the binned light curve published in \cite{Trakhtenbrot}. Originally, F01004-2237 had undergone a luminous flare \citep{f01} in 2010 and later \cite{Trakhtenbrot} classified it as a BFF. The binned light curve of F01004-2237 published in \cite{AT2021loi} shows a bump around a year after its peak. We investigated the recent activity of this source and found that it significantly re-brightened in $2021$ based on its ATLAS monitoring. The recent flare was more luminous than the first one and declined faster. In addition to these two BFFs, AT2021loi shows a bump when declining, roughly a year after the first flare reached its peak. \cite{AT2021loi} suggests that optical re-brightening or bumps seen when decaying might be common features of BFFs. OGLE17aaj, has not shown a clear optical re-brightening, however a bump likely appears roughly a year after the peak, similar to AT2017bgt and AT2021loi. The BFFs classified so far tend to have remarkable variances in their optical light curve evolution despite the similarities. These might be explained by different subclasses of BFFs, similarly to regular TDEs, where different subclasses were defined based on their light curve evolution \citep[e.g.,][]{2023A&A...673A..95C}.

\subsection{Radio characteristics}
\subsubsection{Possible explanations of the long-term radio flare}
\label{subsec:radio_disc}
In this subsection we discuss the long-term radio variability of AT2019aalc. The radio source experienced a brightening of its flux density by a factor of two on a timescale of less than $1.5$\,years, followed by a long-term plateau ($\geq 3$\,years) indicated by the archival radio survey data and our ATCA monitoring. Although radio variability on yearly timescales is a common phenomenon of Seyfert galaxies, the sharp, high-amplitude radio flare and especially the long-term plateau are generally not seen in intrinsically variable Seyfert galaxies \citep[e.g.,][]{seyf_radio_var,2016MNRAS.460..304K} and therefore require further investigation. The unusual radio spectrum with a high frequency turnover is also discussed below.

\begin{itemize}

    \item The brightness temperature of $T_{\mathrm{b}} = (3.0 \pm 0.8) \times 10^{8}$\,K at $1.7$\,GHz significantly exceeds $10^5$\,K, known as an upper limit for supernova remnants and He\,{\sc ii} regions in star-forming galaxies \citep{sfrTb}. Consequently, the radio emission we detected must be non-thermal and originate from physical processes associated to AGN-like (including TDEs) activity (see \citealp{2012MNRAS.426..588B} as an example of how the estimated brightness temperatures help to characterize the radio emission of Seyfert galaxies). This way, we can also safely rule out thermal free-free emission from the accretion disk or the dusty torus \citep[e.g., the case of NGC 1068,][]{1997Natur.388..852G} as the origin of the detected radio emission.

    \item Advection-dominated accretion flows have also been proposed to explain the origin of the radio emission of low-luminosity AGN \citep[e.g., Seyfert galaxies,][]{1997ApJ...477..585M}. Assuming that accretion flows dominate the radio emission of AT2019aalc, following \cite{1998ApJ...499..198Y} we estimate an expected $5$\,GHz radio luminosity of $L_{5\textrm{GHz}}\approx 4 \times 10^{35}$\,erg s$^{-1}$, $3$ orders of magnitude lower than the observed one, clearly indicating that these cannot explain the observed radio emission. 
    
    \item Magnetized coronal winds originating from the AGN accretion disk could produce the non-thermal radio emission of AT2019aalc. \cite{2008MNRAS.390..847L} found a correlation of $\text{log}(L_{R}/L_{X}) \approx -5$ studying the Palomar–Green \citep[PG:][]{1986ApJS...61..305G} radio-quiet quasar sample (in good agreement with the correlation found for coronally active stars by \cite{1993ApJ...405L..63G}). Later, \cite{2015MNRAS.451..517B} came to the same conclusion using a sample of seven radio-quiet Seyfert galaxies. For our source, based on the mean ATCA C-band and the extrapolated \textit{Swift}/XRT $2-20$\,keV luminosities, we estimate a ratio of $\text{log}(L_{R}/L_{X}) \approx -2$, suggesting that magnetized coronal winds alone cannot fully explain the observed radio emission of AT2019aalc. Moreover, coronal emission should be highly variable at radio wavelengths. 

    \item Out-flowing synchrotron-emitting material induced by enhanced accretion is commonly detected in AGN and TDEs in the forms of non-relativistic outflows or relativistic jets \citep[e.g.,][]{2017IAUS..324..119P,2022ApJ...927...74M,2024Galax..12....8K}. The high brightness temperature of the EVN-detected radio feature indicates its radio core identification. The steep ($\alpha<<-0.5$) radio spectrum indicates the presence of another, steep-spectrum synchrotron-emitting component(s) in the system, as radio cores are characterized by a flat radio spectrum. The spectrum may be the composition of the flat-spectrum radio core and the additional, steep-spectrum component(s). The high-frequency turnover around $9$\,GHz supports the idea of a newly ejected radio-emitting component.

\end{itemize}

\subsubsection{The radio spectrum}
The turnover suggests the multi-component nature of the spectrum. Radio spectra with high-frequency excesses have been observed in a few radio-quiet narrow-line Seyfert 1 galaxies \citep{2020A&A...636A..64B}. These sources are thought to be kinematically young AGN with small linear sizes or heavily absorbed AGN. The high-frequency peak might be explained with the kinematic age of a newly ejected component. These extremely young radio sources therefore peak at very high frequencies \citep[e.g.,][]{2020A&A...636A..64B}. The low-frequency peak is more likely originating from star formation activity instead of AGN emission, since the latter is absorbed at these frequencies due to synchrotron self-absorption (SSA) or free-free absorption (FFA).

However, in the case of AT2019aalc, the large-amplitude variability concluded from the VLASS observations and the EVN detection of a high brightness temperature ($T_{b} >> 10^{5}$\,K) feature at $1.7$\,GHz clearly imply AGN-like activity over a star formation scenario at lower frequencies. Therefore, our results indicate that the radio spectrum of AT2019aalc is strongly dominated by AGN activity below $10$\,GHz as well, even if a contribution from star formation, especially at $2.1$\,GHz, is probable. 

A newly ejected component can explain the high-frequency excess, whilst the low-frequency part originates from optically thin synchrotron emission of the AGN. Thus, the overall spectrum represents a superposition of past and recent AGN activity, indicative of its recently enhanced, intermittent behavior. \cite{2021FrASS...8..147J} states that restarted AGN activity can explain at least some of the radio spectra studied by \cite{2020A&A...636A..64B}. Intermittent activity is further supported by theoretical models highlighting that sources with high accretion rates are more prone to this kind of behavior \citep{2009ApJ...698..840C}.

\subsection{X-ray properties}
The X-ray spectra of AGN at energies above $2$\,keV have a power-law-like shape. These X-ray photons are believed to be triggered by Compton up-scattering of the accretion disk photons off hot electrons surrounding the disk, in a hot ($\approx 10^{9}$\,K) optically thin corona above the disk \citep[e.g.,][]{2015A&A...582A..40K}. Below $2$\,keV some AGN show an excess in their spectrum referred to as the soft X-ray excess. It can be modeled well by a blackbody model with a best-fit temperature in the range $0.1-0.2$\,keV. The origin of the excess might be explained via Comptonized disk emission. The reflection of hard X-ray photons from the surface of the disk can also account for the excess \citep{2012MNRAS.420.1848D}.

The X-ray spectrum of AT2019aalc exhibits a non-thermal component with a power-law index of $\Gamma \approx 3.3$ and an unusually soft thermal component with a blackbody temperature of $kT \approx 95$\,eV. Most of the BLSy1s typically have a harder X-ray spectrum with a power-law index of $\Gamma = 1.7-2$ \citep[e.g.,][]{2014A&A...568A.108P}. AT2019aalc exhibits properties that are more common in sources characterized by enhanced accretion, e.g., NLSy1s or TDEs which both have typically soft X-ray spectra with $\Gamma > 2$ \citep{1996A&A...305...53B,2020SSRv..216...85S}.

The soft X-ray flux of Seyfert galaxies was found to be correlated well with the observed fluxes in the coronal line [Fe\,{\sc x}] $\lambda 6375$. The coronal lines to X-ray flux ratio is log$(f_{\mathrm{[Fex]}}/f_{x}) = -3.43 \pm 0.55$ for both broad and narrow-line Seyfert 1 galaxies \citep{2009MNRAS.397..172G}. For AT2019aalc we estimate values of log$(f_{\mathrm{[Fex]}}/f_{x}) \approx -3.45$ when optical spectroscopic and X-ray observations were performed nearly simultaneously (due to the rapid X-ray variability, we consider only observations taken on the same day). This suggests that this coronal line is excited by the X-ray flares discussed here instead of past X-ray flares that are thought to power the ECLEs.

The reoccurring optical/UV bumps appearing during the decay of the main optical flares are likely driven by the soft X-ray flares, which lead these outbursts only by a couple of weeks. The soft X-ray excess appears to be more dominant during these flares. Interestingly, the soft X-ray flares show hints of quasi-periodicity of $\approx 90$\,days. A few Repeating Nuclear Transients (RNTs) were discovered in the past few years: Quasi-Periodic Eruptions (QPEs) exhibit soft X-ray flares on time-scales of hours to days, while partial TDEs exhibit such flares on timescales of months to years \citep[e.g.,][]{qpe2,qpe1,qpe3,qizqpe}. We note that another BFF, AT2020afhd, shows hints of quasi-periodic soft X-ray flares according to its \textit{Swift}/XRT monitoring. This source is included  in the The Living Swift XRT Point Source Catalogue (LSXPS: \citealp{2023MNRAS.518..174E}) as LSXPS J031335.6$-$020905. The potential connection between BFFs and QPEs should be explored in future studies.

\subsection{SED fitting}
\begin{figure}
  \centering
  {\includegraphics[width=100mm]{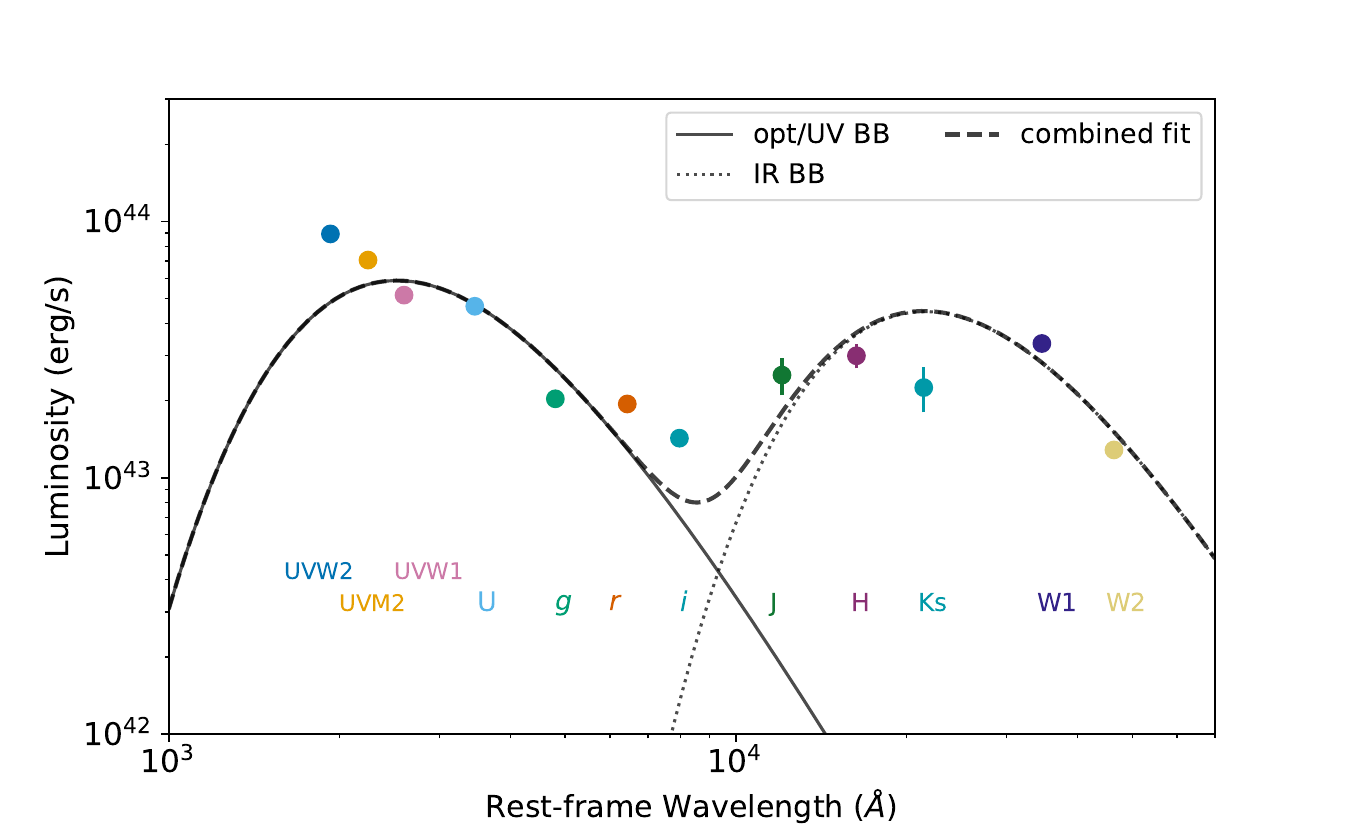}}
  \caption{The extinction-corrected and host-subtracted SED of the second flare of AT2019aalc as measured around the optical peak. Two blackbody models to the optical/UV (solid curve) and IR data points (dotted curve) were fitted separately. A combined fit (black dashed curve) is also shown.}
  \label{pic:bbody}
\end{figure}

Following \cite{Tywin}, we fitted two blackbody models in order to describe the SED of the transient using \textsc{lmfit}. These curves fitted to the optical/UV (blue blackbody) and IR datapoints (red blackbody) are shown in Fig. \ref{pic:bbody}, together with a combined double blackbody fit. The red blackbody fit was introduced in order to account for the IR dust echo. These curves fit the SED around the optical peak of the investigated flare.

The continuum emission of TDEs is described well by a thermal blackbody model \citep{2021ARA&A..59...21G}. The case of AT2019aalc appears to be more complex. The UV emission has a contribution from the Bowen Fluorescence mechanism and possibly also from Comptonization, explaining the poor fit towards the blue. The excess seen in the WISE bands is possibly due to the non-simultaneous observation with respect to the optical/UV and WIRC observations, as the IR light curve is peaked well after the optical one. The excess in the $r$-band possibly has a contribution from the early appearance of some Fe lines in this range, but such a significant excess cannot be fully explained with the coronal lines only. Instead, a variable H$\alpha$ component, given its much larger equivalent width, or reprocessing, is more likely to significantly contribute to the observed excess and warrants consideration.

The inferred blackbody temperature and radius values estimated for the blue blackbody fit are $T_{\textrm{blue}}\approx14000$\,K and $R_{\textrm{blue}}\approx1.5 \times 10^{14}$\,cm, respectively. The blackbody temperature is in the lower range of that is found for optical selected TDEs, however, an order of magnitude lower than the typical values of the X-ray detected TDEs studied by \cite{2020SSRv..216..124V} and \cite{2022ApJ...937L..28T}. On the other hand, the radius is smaller than those calculated for optically selected TDEs, however is more consistent with the X-ray detected ones. The red blackbody fit resulted in $T_{\textrm{red}}\approx1400$\,K and $R_{\textrm{red}}\approx2.8 \times 10^{16}$\,cm.

\subsection{What type of transient is AT2019aalc?}
\label{subsect:classification}
AT2019aalc shares similarities with TDEs including the soft X-ray spectrum, the dust echo and the optical flare(s) rising with constant color. The UV bright nature of the transient with a blue optical/UV peak color of NUV$-r \approx 0.2$ is another TDE-like feature and not typical for AGN and SNe \citep{2021ARA&A..59...21G}. Bowen lines appear in the spectra of several TDEs even if these are typically weaker than in our case. The most remarkable difference is, however, the fading of the optical emission, which is slower than for any TDE identified so far. The reddening in color when the main optical flares are declining is also atypical for TDEs. However, in a more complex AGN environment, compared to the simpler case of standard TDEs in quiescent galaxies, reprocessing may already be significant at early times, implying that the observed reddening does not necessarily reflect genuine cooling.

The unobscured AGN-like optical spectra, the presence of persistent and strong Bowen lines (O\,{\sc iii} $\lambda 3133$ and  N\,{\sc iii} $\lambda 4640$), the extremely luminous UV flare, optical emission decaying on yearly timescales, and the optical re-brightening altogether suggest a Bowen Fluorescence Flare classification of AT2019aalc. Furthermore, we classify AT2019aalc as an Extreme Coronal Line Emitter because of the detection of several strong high-ionization coronal lines.

The mechanism that powers the BFFs is still unclear. TDEs in AGN might play a role. In a single star case, the slowly decaying optical/UV emission might be explained by the dissipation of energy due to the continuous interaction between the stellar debris stream and the pre-existing accretion disk (similarly to the case of the TDE-AGN PS16dtm, \citealp{2017ApJ...843..106B}, and ASASSN-14ko, \citealp{2023ApJ...956L..46H}) or the AGN's radiation field. According to TDE-AGN simulations, stellar passages can induce substantial perturbations in the inner disk and leave it in a very perturbed state even for years \citep{tdes_in_agn}. The debris-disk interaction results in enhanced accretion and multiple shocks \citep{tdes_in_agn}, notably leading to X-ray emission \citep{2024MNRAS.527.8103R}. Consistently within this picture, the soft X-ray flares of AT2019aalc suggest that the emission is linked to the accretion disk. These X-ray flares lead the optical/UV bumps by $1-2$\,weeks, which are detected during the decay of the main flare. The soft X-ray emission is likely reprocessed to optical/UV by the dense environment, explaining the optical/UV flares and the emergence of the Bowen fluorescence and coronal lines. The TDE-AGN scenario provides an explanation for one slowly decaying flare with recurring outbursts, however, AT2019aalc exhibits two distinct optical flares. TDE-AGN simulations also revealed that in a case of an initially partial disruption, a surviving core would generate similar heating of the inner disk to the first pericenter passage \citep{tdes_in_agn,2024MNRAS.527.8103R}. Notably, the second flare of AT2019aalc is more luminous than the first one, similarly to the cases of the partial TDE AT2020vdq \citep{2020vdq} and the BFF F01004-2237. Numerical simulations demonstrated that the even more luminous second flare of a partial TDE might be explained in the following way: tidal energy injection leads to oscillations in the remnant star, and if the thermal energy cannot dissipate within one orbital period, the star expands, making it more susceptible to further tidal disruption and potentially resulting in brighter flares \citep{vdq2,partial_models}. \cite{vdq2} claimed that these considerations provide a plausible explanation for the brighter second flare of the partial TDE AT2020vdq. A partial TDE scenario might explain the second optical flare of the other two BFFs (AT2021loi and F01004-223) with double-peaked optical flares as well.

A TDE-in-AGN scenario would explain the TDE-like properties of these sources. The most remarkable differences, such as the slowly decaying optical emission and the continuum emission not being well described by blackbody models, may arise due to the more complex environment in which the TDE occurs, compared to the simpler conditions in quiescent galaxies ("TDEs in vacuum"). TDE-AGN simulations showed that the emergent spectrum of such a system may not be purely thermal as in regular TDEs \citep{tdes_in_agn}. Furthermore, a possible connection between the high-ionization coronal lines (which are most probably related to TDE activity, \citealp{2012ApJ...749..115W}) and the BFFs (see Fig.~\ref{pic:coronal}) is consistent with this picture. A known candidate partial TDE in an AGN is the case of the active galaxy IC 3599. This low-luminosity AGN exhibits periodic soft X-ray flares interpreted as being due to interaction with a partially disrupted star, which results in reoccurring outbursts after each passage of the surviving core around the central black hole \citep{2015A&A...581A..17C}. Notably, IC 3599 has more prominent high-ionization coronal lines than most AGN (similarly to the coronal-line detected transients, see Fig.~\ref{pic:coronal}). In addition, IC 3599 is the only known AGN with fading coronal lines \citep{2019ApJ...883...31F} a behavior also observed in the TDE AT2019qiz \citep{2023MNRAS.525.1568S}, the BFF F01004-2237 \citep{f01} and most of the ECLEs \citep{2013ApJ...774...46Y}. Another feature AT2019aalc shares with other candidate pTDEs is the He{\sc ii} $\lambda 4686$ + N{\sc iii} $\lambda 4640$ complex, observed in the optical spectra of all five optically selected pTDE candidates (see Fig. 13. in \cite{2024arXiv241215326H}).

Interestingly, other BFFs also showed weaker pre-transient AGN activity than persistently accreting AGN \citep{Trakhtenbrot,AT2021loi}. It suggests that TDEs might have reactivated these AGN. However, between the two main optical flares, AT2019aalc exhibited negative fluxes, i.e., it was fainter than the baseline (see Fig.~\ref{pic:ztf_diff}). This behavior suggests a retrograde orbit of the disrupted star relative to the accretion disk, as such encounters gain angular momentum and lead to a subsequent lower AGN state \citep{2022MNRAS.514.4102M}. In contrast, a prograde orbit adds angular momentum to the disk, resulting in a higher AGN state, which may explain the light curve of OGLE17aaj, where the source persisted in a high and stable state for years after the peak \citep{2019A&A...622L...2G}, unlike AT2019aalc.

\subsection{Radio properties of BFFs}
As we studied AT2019aalc in radio in detail, we compare the
radio properties with that of the BFFs. Based on the radio observations explained in \cite{AT2021loi}, AT2021loi has an inverted radio spectrum between $5$ and $10$\,GHz while the source was not detected at $3$\,GHz. Although no radio spectrum is available for AT2017bgt, we investigated the VLASS QL Images of this BFF to study its radio flux variability. We found a radio source at the optical position of AT2017bgt with flux densities of $3.6$\,mJy (on 2019-05-21), $6.6$\,mJy (on 2021-10-19) and $4.5$\,mJy (on 2024-06-11) at a central frequency of $3$\,GHz. The pre-flare system was detected in the FIRST (at $1.4$\,GHz) in July 1998 with a flux density of $0.9$\,mJy. F01004-2237 was detected with VLBI at $8.4$\,Ghz, however, remained undetected at $2.3$\,GHz which indicates an inverted spectrum \citep{2024ApJ...970....5H}. The BFF AT2020afhd was detected with the VLA at $15$\,GHz with $0.25$\,mJy \citep{2024TNSAN..56....1C}. If BFFs are explainable with TDE-AGN, the synchrotron emission is likely originating from outflow activity due to the enhanced accretion or, alternatively, from interaction between the unbound debris and the surrounding gas predicted by TDE-AGN simulations \citep{tdes_in_agn}. Radio observations of any newly discovered BFF will be essential to better understand these transients.

\subsection{BFFs and high-energy neutrinos}
\begin{figure}
  \centering
  {\includegraphics[width=90mm]{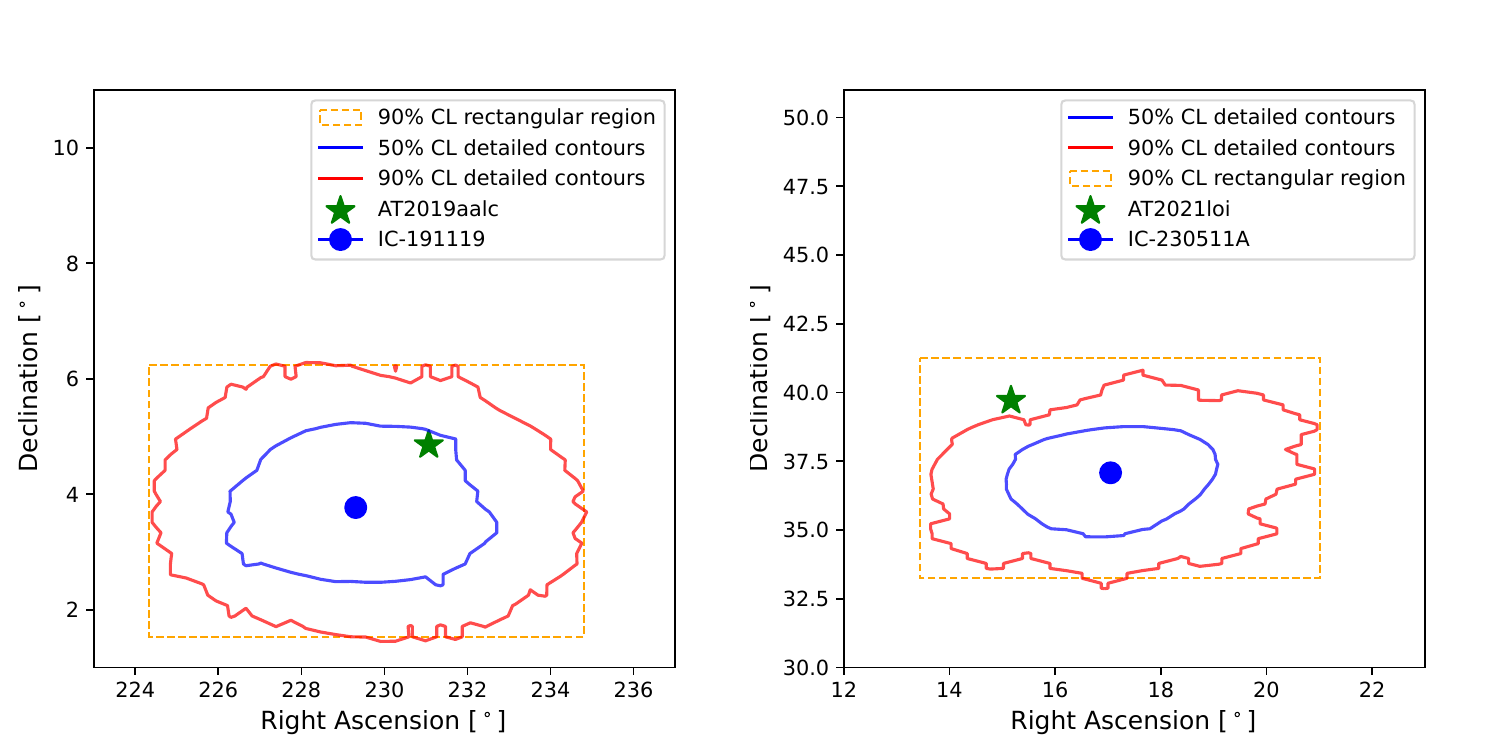}}
  \caption{The blue and red contours represent the 50\% and 90\%CL containment detailed contours of the high-energy neutrino events IC191119A (left) and IC230511 (right), respectively. The orange box is the 90\%CL rectangular containment region. The neutrino best-fit position is shown with a blue point. The optical positions of AT2019aalc (left) and AT2021loi (right) are marked with green star signs.}
  \label{pic:neu}
\end{figure}

AT2019aalc was associated with the high-energy neutrino event IC191119A (most probable neutrino energy: $177$\,TeV) detected by the IceCube Observatory \citep{2019GCN.26258....1I,Lancel}. Motivated by the classification of AT2019aalc suggested by its multi-wavelength properties (Subsect.~\ref{subsect:classification}), we searched for spatial coincidences between the earlier-classified BFFs and the neutrino alerts from the IceCube Event Catalog of Alert Tracks \citep[ICECAT-1:][]{2023ApJS..269...25A}. We found that the BFF AT2021loi is located within the 90\% rectangular uncertainty contour
of the high-energy neutrino event IC230511 (most probable neutrino energy: $167$\,TeV) detected in May $2023$ \citep{2023GCN.33773....1I}. In the full likelihood contours published in ICECAT-1, the source is just outside the 90\% contour. We present the source positions together with the neutrino best-fit positions and both contours in Fig~\ref{pic:neu}. We also report a coincidence between the prototype BFF AT2017bgt and the high-energy neutrino event IC-200410A. This event is, however, a poorly reconstructed cascade event, therefore we do not discuss it further here.

The neutrino associated with AT2021loi was detected $680$\,days and $290$\,days after the first and second optical peaks, respectively. Notably, all three nuclear transients (including AT2019aalc) associated with neutrinos exhibited a similar pattern, with neutrino detection occurring $150-290$\,days after the optical peaks. This might be explained with delayed outflow activity \citep{Lancel}. In the case of AT2021loi, the inverted radio spectrum \citep{AT2021loi} is compatible with a newly ejected outflow component, similarly to the case of AT2019aalc.  Interactions between the out-flowing material and clouds in the surrounding of the SMBH \citep{2022MNRAS.514.4406W} or between ultra-high-energy protons (accelerated in a jet or outflow) and infrared photons from the dust echo \citep{2023ApJ...948...42W} were proposed to explain the observed neutrino emission of the three nuclear transients. These findings indicate that outflows play an important role in the neutrino production. A case study for AT2021loi would help us to further constrain the neutrino production models of nuclear transient sources.echos

If the connection between the high-ionization coronal line detection and the unusually large dust echos of nuclear transients is real, this can be used to reveal more sources with strong dust echoes and therefore more candidate neutrino sources. Notably, both AT2019aalc and AT2021loi are characterized by significant dust echoes (\citealp{Lancel,2024MNRAS.531.2603H}: see also Subsect.~\ref{subsect:ir_res}) and unusually strong high-ionization coronal lines (\citealp{AT2021loi}: see also Subsect.~\ref{subsect:spectro}).

\section{Conclusions}
\label{sect_6}
We conducted a multi-wavelength campaign during the re-brightening of the nuclear transient AT2019aalc, which have been associated with a high-energy neutrino event. We found that the second optical flare was more luminous and decayed slower than the first one. Both flares rose rapidly and with constant color. The UV emission peaked at extremely high luminosities (at $\approx10^{44}$\,erg/s). The re-brightening was accompanied by an IR flare, similarly to the first flare, when it was explained with a dust echo emission. An X-ray flare peaked before the optical/UV emission, preceding two subsequent flares that were nearly an order of magnitude brighter and predominantly blackbody in nature. The X-ray flares are very soft, which explains the unusually strong He\,{\sc ii} lines. In the optical spectra, persistent Bowen fluorescence lines were detected over a year after the second flare. With VLBI, we clearly detect a non-thermal radio source, which favors an AGN-like origin. The radio flux monitoring revealed a long-term flare and an inverted spectrum above $9$\,GHz, which is likely explained by a newly ejected, self synchrotron-absorbed outflow component. Altogether, our results suggest a Bowen Fluorescence Flare classification of AT2019aalc. We also classify AT2019aalc as an Extreme Coronal Line Emitter. 

The two distinct optical flares might be attributed to the partial disruption of a single star within an AGN environment. The stellar debris-disk interaction in such a system could result in complex light curve variations, e.g., the plateau observed in optical/UV \citep{tdes_in_agn_2}. A TDE-related scenario provides an explanation for the suddenly enhanced activity of the nucleus. Future observations will help us to better understand the nature of AT2019aalc: a repeating TDE would lead to a roughly periodic flaring, while some other AGN activity would lead to erratic future flares. 

Motivated by the high-energy neutrino associated with AT2019aalc, we cross-matched the ICECAT-1 with the positions of the four known BFFs. We found that AT2021loi is coincident with the high-energy neutrino event IC230511, confirming the flaring AGN-neutrino connection studied by \cite{Lancel}. Apart from the known multi-wavelength properties of BFFs, AT2021loi shares characteristics with AT2019aalc such as late-time IR flare (due to dust echo), high-ionization coronal lines in its spectra (and also $r$-band excess), radio detection (and inverted radio spectrum) and a re-brightening episode in optical. In both cases the AGN are classified as broad-line AGN prior to when the transients occurred.

\begin{acknowledgements}
We sincerely thank the anonymous referees for their detailed comments and insightful suggestions, which have greatly enhanced the clarity and quality of this paper. We thank Benny Trakhtenbrot for useful discussions. We thank Elisa Pueschel for proofreading. PMV, AF, BA, ST, AK, DJB, EM-B acknowledge the support from the DFG via the Collaborative Research Center SFB1491 \textit{Cosmic Interacting Matters - From Source to Signal}. EH acknowledges support by NASA under award number 80GSFC21M0002. We thank the \textit{Swift} team for approving and conducting our ToO requests.
\newline
The European VLBI Network is a joint facility of independent European, African, Asian, and North American radio astronomy institutes. Scientific results from data presented in this publication are derived from the following EVN project code: EV027. e-MERLIN is a National Facility operated by the University of Manchester at Jodrell Bank Observatory on behalf of STFC. We acknowledge the use of public data from the Swift data archive. This research has made use of the XRT Data Analysis Software (XRTDAS) developed under the responsibility of the ASI Science Data Center (ASDC), Italy. The Australia Telescope Compact Array is part of the Australia Telescope National Facility (\url{https://ror.org/05qajvd42}) which is funded by the Australian Government for operation as a National Facility managed by CSIRO. We acknowledge the Gomeroi people as the Traditional Owners of the Observatory site. Funding for the Sloan Digital Sky Survey V has been provided by the Alfred P. Sloan Foundation, the Heising-Simons Foundation, the National Science Foundation, and the Participating Institutions. SDSS acknowledges support and resources from the Center for High-Performance Computing at the University of Utah. SDSS telescopes are located at Apache Point Observatory, funded by the Astrophysical Research Consortium and operated by New Mexico State University, and at Las Campanas Observatory, operated by the Carnegie Institution for Science. The SDSS web site is \url{www.sdss.org}. SDSS is managed by the Astrophysical Research Consortium for the Participating Institutions of the SDSS Collaboration, including Caltech, The Carnegie Institution for Science, Chilean National Time Allocation Committee (CNTAC) ratified researchers, The Flatiron Institute, the Gotham Participation Group, Harvard University, Heidelberg University, The Johns Hopkins University, L’Ecole polytechnique f\'{e}d\'{e}rale de Lausanne (EPFL), Leibniz-Institut für Astrophysik Potsdam (AIP), Max-Planck-Institut für Astronomie (MPIA Heidelberg), Max-Planck-Institut für Extraterrestrische Physik (MPE), Nanjing University, National Astronomical Observatories of China (NAOC), New Mexico State University, The Ohio State University, Pennsylvania State University, Smithsonian Astrophysical Observatory, Space Telescope Science Institute (STScI), the Stellar Astrophysics Participation Group, Universidad Nacional Aut\'{o}noma de M\'{e}xico, University of Arizona, University of Colorado Boulder, University of Illinois at Urbana-Champaign, University of Toronto, University of Utah, University of Virginia, Yale University, and Yunnan University. This publication makes use of data products from the Wide-field Infrared Survey Explorer, which is a joint project of the University of California, Los Angeles, and the Jet Propulsion Laboratory/California Institute of Technology, and NEOWISE, which is a project of the Jet Propulsion Laboratory/California Institute of Technology. WISE and NEOWISE are funded by the National Aeronautics and Space Administration. The National Radio Astronomy Observatory is a facility of the National Science Foundation operated under cooperative agreement by Associated Universities, Inc. This work has made use of data from the Asteroid Terrestrial-impact Last Alert System (ATLAS) project. The Asteroid Terrestrial-impact Last Alert System (ATLAS) project is primarily funded to search for near earth asteroids through NASA grants NN12AR55G, 80NSSC18K0284, and 80NSSC18K1575; byproducts of the NEO search include images and catalogs from the survey area. This work was partially funded by Kepler/K2 grant J1944/80NSSC19K0112 and HST GO-15889, and STFC grants ST/T000198/1 and ST/S006109/1. The ATLAS science products have been made possible through the contributions of the University of Hawaii Institute for Astronomy, the Queen’s University Belfast, the Space Telescope Science Institute, the South African Astronomical Observatory, and The Millennium Institute of Astrophysics (MAS), Chile. This work is based on data from eROSITA, the soft X-ray instrument aboard SRG, a joint Russian-German science mission supported by the Russian Space Agency (Roskosmos), in the interests of the Russian Academy of Sciences represented by its Space Research Institute (IKI), and the Deutsches Zentrum für Luft- und Raumfahrt (DLR). The SRG spacecraft was built by Lavochkin Association (NPOL) and its subcontractors, and is operated by NPOL with support from the Max Planck Institute for Extraterrestrial Physics (MPE). The development and construction of the eROSITA X-ray instrument was led by MPE, with contributions from the Dr. Karl Remeis Observatory Bamberg \& ECAP (FAU Erlangen-Nuernberg), the University of Hamburg Observatory, the Leibniz Institute for Astrophysics Potsdam (AIP), and the Institute for Astronomy and Astrophysics of the University of Tübingen, with the support of DLR and the Max Planck Society. The Argelander Institute for Astronomy of the University of Bonn and the Ludwig Maximilians Universität Munich also participated in the science preparation for eROSITA. The Pan-STARRS1 Surveys (PS1) and the PS1 public science archive have been made possible through contributions by the Institute for Astronomy, the University of Hawaii, the Pan-STARRS Project Office, the Max-Planck Society and its participating institutes, the Max Planck Institute for Astronomy, Heidelberg and the Max Planck Institute for Extraterrestrial Physics, Garching, The Johns Hopkins University, Durham University, the University of Edinburgh, the Queen's University Belfast, the Harvard-Smithsonian Center for Astrophysics, the Las Cumbres Observatory Global Telescope Network Incorporated, the National Central University of Taiwan, the Space Telescope Science Institute, the National Aeronautics and Space Administration under Grant No. NNX08AR22G issued through the Planetary Science Division of the NASA Science Mission Directorate, the National Science Foundation Grant No. AST-1238877, the University of Maryland, Eotvos Lorand University (ELTE), the Los Alamos National Laboratory, and the Gordon and Betty Moore Foundation.  These results made use of the Lowell Discovery Telescope (LDT) at Lowell Observatory.  Lowell is a private, non-profit institution dedicated to astrophysical research and public appreciation of astronomy and operates the LDT in partnership with Boston University, the University of Maryland, the University of Toledo, Northern Arizona University and Yale University. The upgrade of the DeVeny optical spectrograph has been funded by a generous grant from John and Ginger Giovale and by a grant from the Mt. Cuba Astronomical Foundation. Some of the data presented herein were obtained at Keck Observatory, which is a private 501(c)3 non-profit organization operated as a scientific partnership among the California Institute of Technology, the University of California, and the National Aeronautics and Space Administration. The Observatory was made possible by the generous financial support of the W. M. Keck Foundation. The authors wish to recognize and acknowledge the very significant cultural role and reverence that the summit of Maunakea has always had within the Native Hawaiian community. We are most fortunate to have the opportunity to conduct observations from this mountain. This publication makes use of data products from the Two Micron All Sky Survey, which is a joint project of the University of Massachusetts and the Infrared Processing and Analysis Center/California Institute of Technology, funded by the National Aeronautics and Space Administration and the National Science Foundation. Based on observations obtained with the Samuel Oschin Telescope 48-inch and the 60-inch Telescope at the Palomar Observatory as part of the Zwicky Transient Facility project. ZTF is supported by the National Science  Foundation under Grant No. AST-2034437 and a collaboration including Caltech, IPAC, the Weizmann Institute for Science, the Oskar Klein Center at Stockholm University, the University of Maryland, Deutsches Elektronen-Synchrotron and Humboldt University, the TANGO Consortium of Taiwan, the University of Wisconsin at Milwaukee, Trinity College Dublin, Lawrence Livermore National Laboratories, and IN2P3, France. Operations are conducted by COO, IPAC, and UW.

\end{acknowledgements}

\bibliographystyle{aa}
\bibliography{biblio}

\onecolumn
\appendix
\section{Spetroscopic details}
\begin{table*}[h]
    \begin{center}
        \caption{Summary of the optical spectroscopic observations of AT2019aalc. $\delta t_{f2}$ is time relative to the optical peak of the second flare.}
        \label{table:sp}
        \begin{tabular}{cccccc}
            \toprule 
            Date & Telescope & Instrument & Exp. time & Slit width & Seeing FWHM\\
            & & \\
            & & & (s) & (arcsec) & (arcsec) \\
            \hline
            \hline
            &&Post-flare-1&& \\
            \hline
            2021-07-06 & Keck-I & LRIS & 300 & 1.0 & 1.1 \\
            \hline
            &&Over-flare-2&& \\
            \hline
            2023-06-15 & Keck-I & LRIS & 300 & 1.0 & 1.4 \\
            2023-06-22 & LDT & DeVeny &  800 & 1.5 & 1.2 \\
            2023-07-11 & Palomar Hale $5$\,m & DBSP & 180 & 1.5 & 1.5  \\
            2023-08-17 & Palomar Hale $5$\,m & DBSP & 900 & 1.5 & 1.1 \\
            2024-05-21 & Palomar Hale $5$\,m & DBSP & 1200 & 1.5 & 1.3 \\
            \bottomrule
        \end{tabular}
    \end{center}
\end{table*}

\begin{table*}[h]
\centering
\caption{Emission line properties of AT2019aalc resulted from the fitting procedure. $\delta$t is relative to the optical peak. \textit{F}(BF) represents the flux of the N\,{\sc iii} $\lambda 4640+$He\,{\sc ii} $\lambda 4686$ complex. $n_\textrm{coronal}$ gives the number of the detected high-ionization coronal lines. Superscripts specify which lines are detected, with their labels corresponding to: (a) [Fe\,{\sc vii}]$\lambda 6087$, (b) [Fe\,{\sc x}]$\lambda 6375$, (c) [Fe\,{\sc xiv}]$\lambda 5303$, (d) [Fe\,{\sc xii}]$\lambda 5721$ and (e) [Fe\,{\sc xi}]$\lambda 7892$.}\label{table:a2}
\begin{tabular}{cccccc}
\toprule
Date & Instrument & Resolving power & $\delta t_{f2}$ & $F(\mathrm{BF}) / F(\mathrm{H}\beta)$ & $n_\textrm{coronal}$ \\
& & (approx.) & (days) & & \\
\hline
2008-04-05 & SDSS & 1500-2500 & -11 years & & 2$^{a,b}$\\
\hline
2021-06-07 & LRIS  & 1000 & -737 & 0.21 $\pm$ 0.01 & 4$^{a,b,c,d}$ \\
2023-06-15 & LRIS & 1000 & -28 & 0.64 $\pm$ 0.02 & 4$^{a,b,c,d}$\\
2023-06-22 & Deveny & 750 & -21 & 0.50 $\pm$ 0.01 & 4$^{a,b,c,d}$ \\
2023-07-11 & DBSP  & 1000 & -2 & 0.83 $\pm$ 0.04 & 4$^{a,b,c,d}$ \\
2023-08-17 & DBSP  & 1000 & +35 & 0.25 $\pm$ 0.01 & 5$^{a,b,c,d,e}$ \\
2024-05-21 & DBSP  & 1000 & +315 & 0.30 $\pm$ 0.02 & 5$^{a,b,c,d,e}$ \\ 
\bottomrule
\end{tabular}
\end{table*}

\begin{figure*}[!htbp]
  \centering
  {\includegraphics[width=0.87\textwidth]{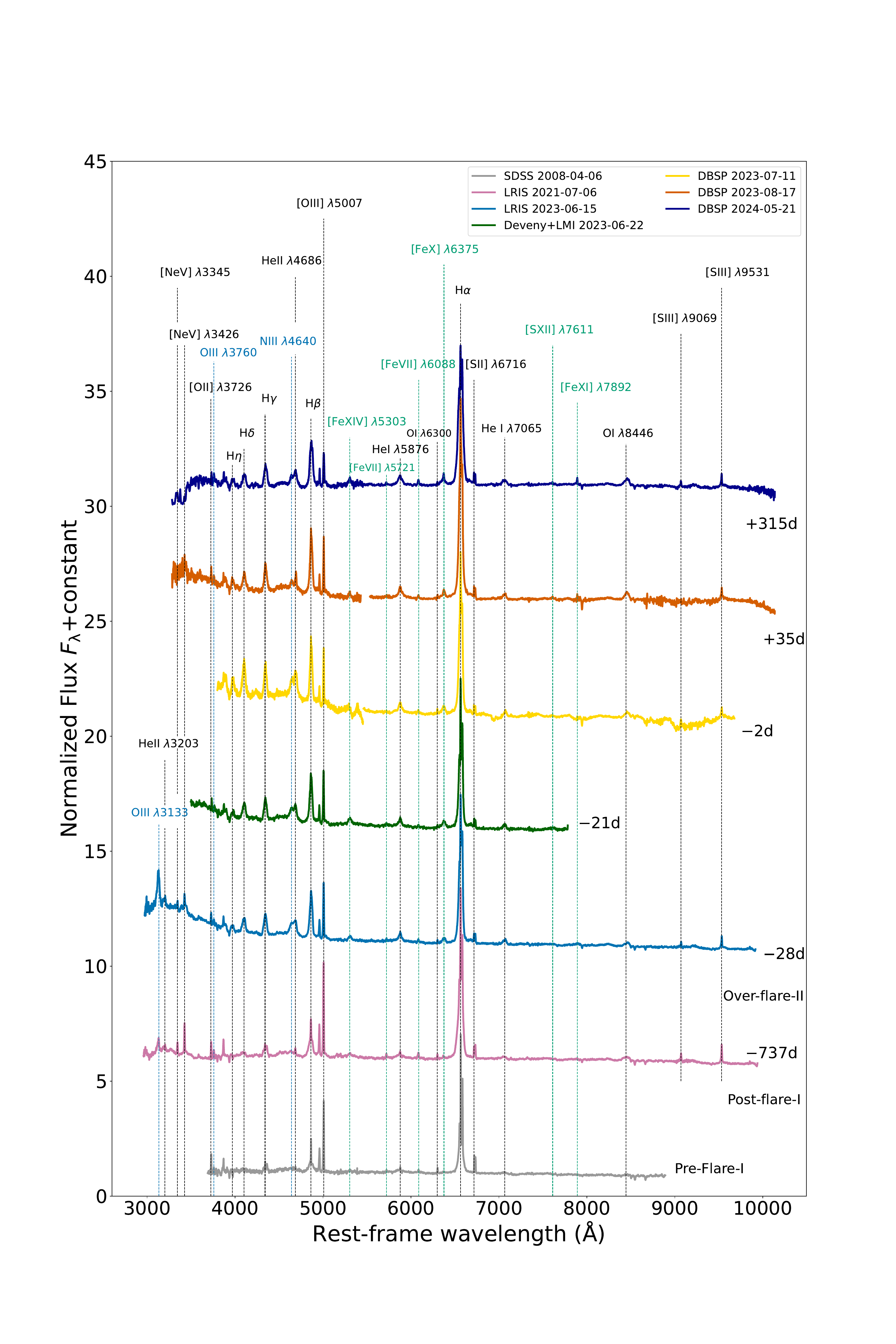}}
  \caption{The raw optical spectra of AT2019aalc. The remarkable lines are indicated with dotted vertical lines. Days relative to the second optical peak are given to the right from each spectra. The SDSS pre-flare spectrum was observed around $11$\,years before the first optical flare.}
  \label{fig:a1}
\end{figure*}

\begin{figure*}[h]
  \centering
 \includegraphics[width=90mm]{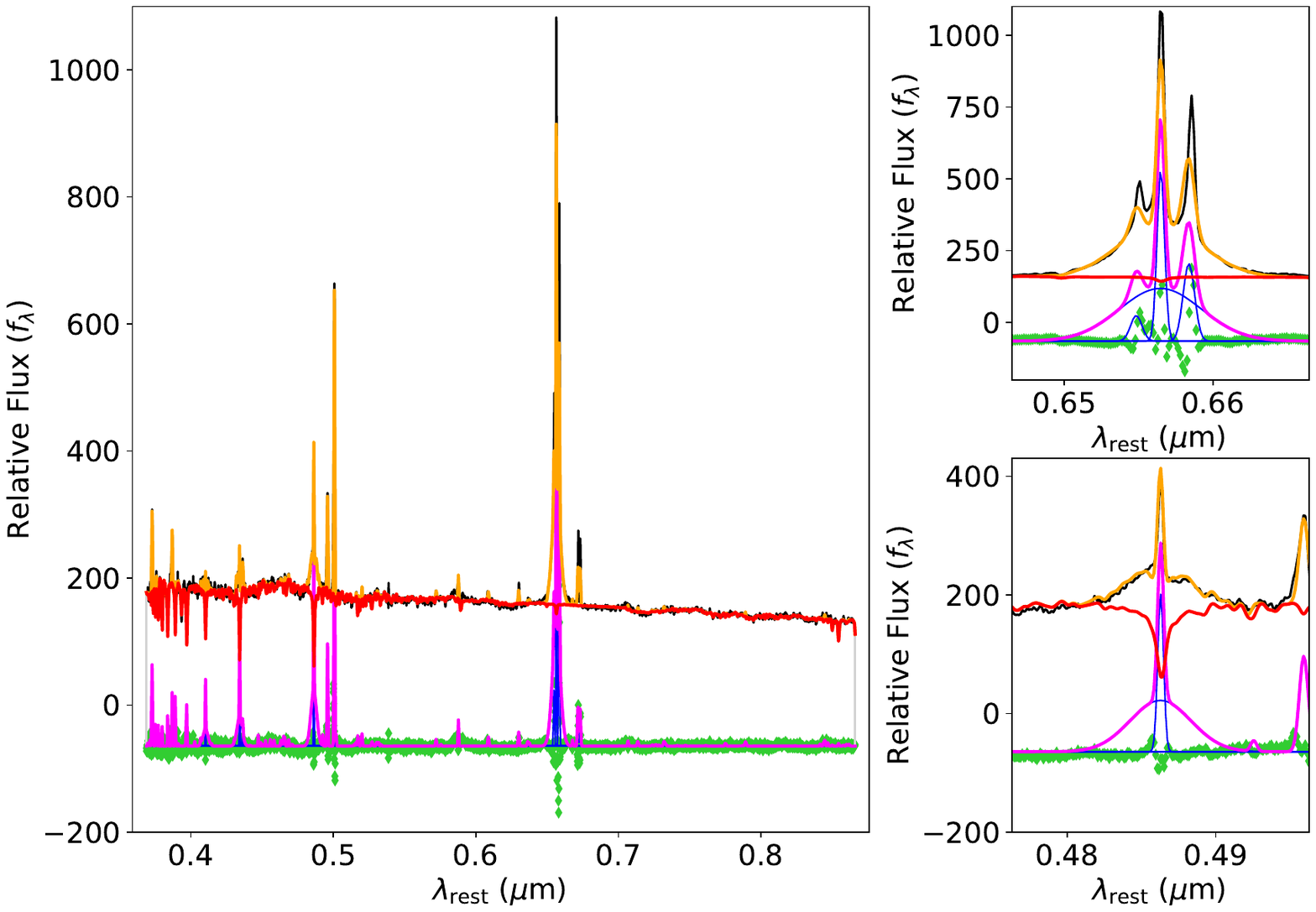}
  \caption{The pre-flare SDSS spectrum provided by \textsc{pPXF} supplemented by two subplots showing $\mathrm{H}\alpha$ (top right corner) and $\mathrm{H}\beta$ (lower right corner). The input spectrum is shown as solid black line, the best stellar and stellar + gas fit are represented as solid red and orange lines, respectively \citep{2023MNRAS.526.3273C}. Additionally, plotted with an arbitrarily chosen offset, one can see the individual gas components (solid blue), total gas emission (solid magenta) and fit residuals (green diamonds) \citep{2023MNRAS.526.3273C}.}
  \label{fig:a2}
\end{figure*}

\label{app}
\begin{figure*}[h]
  \centering
  \includegraphics[width=0.45\textwidth]{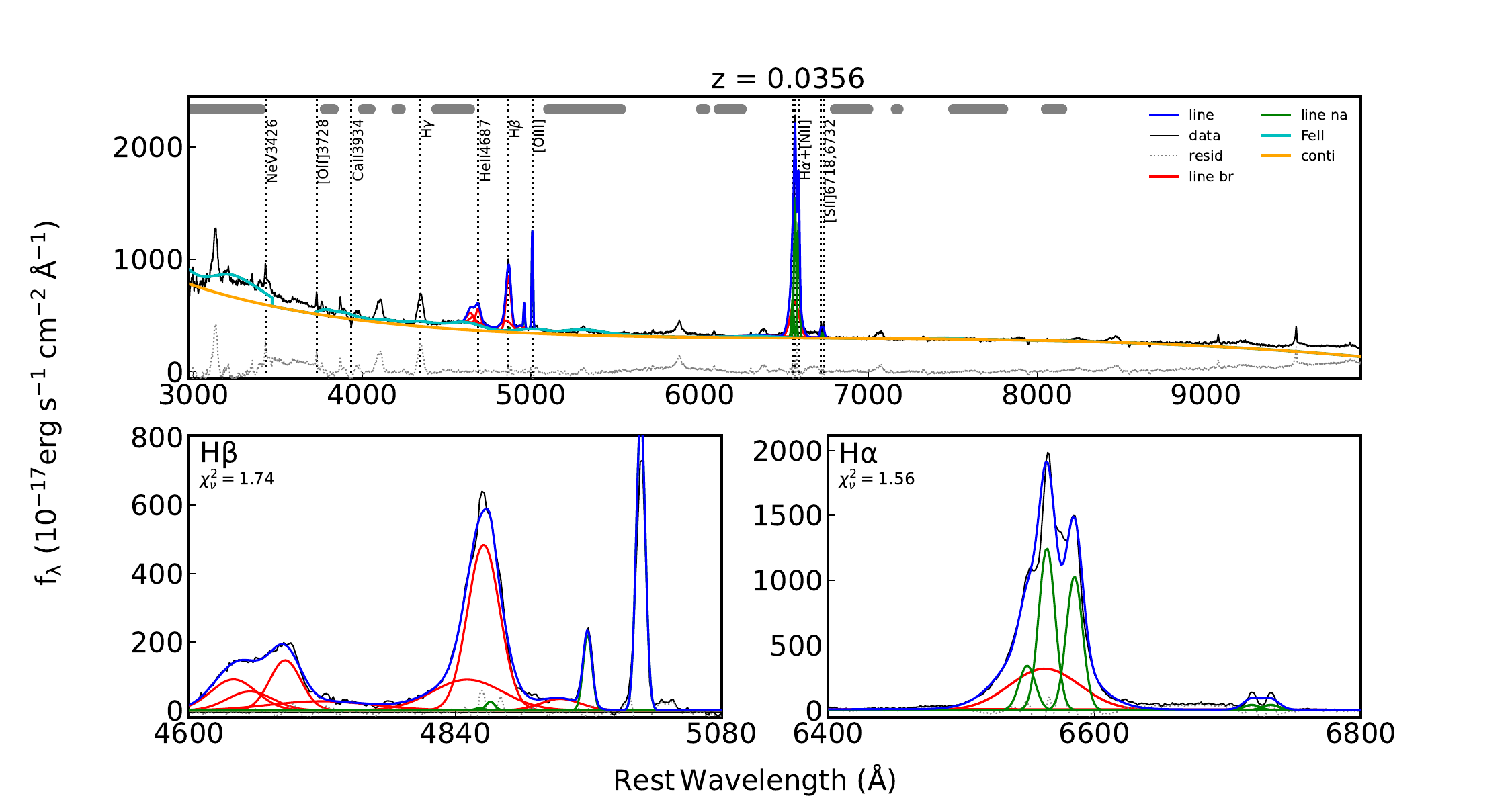}
  \hfill
  \includegraphics[width=0.45\textwidth]{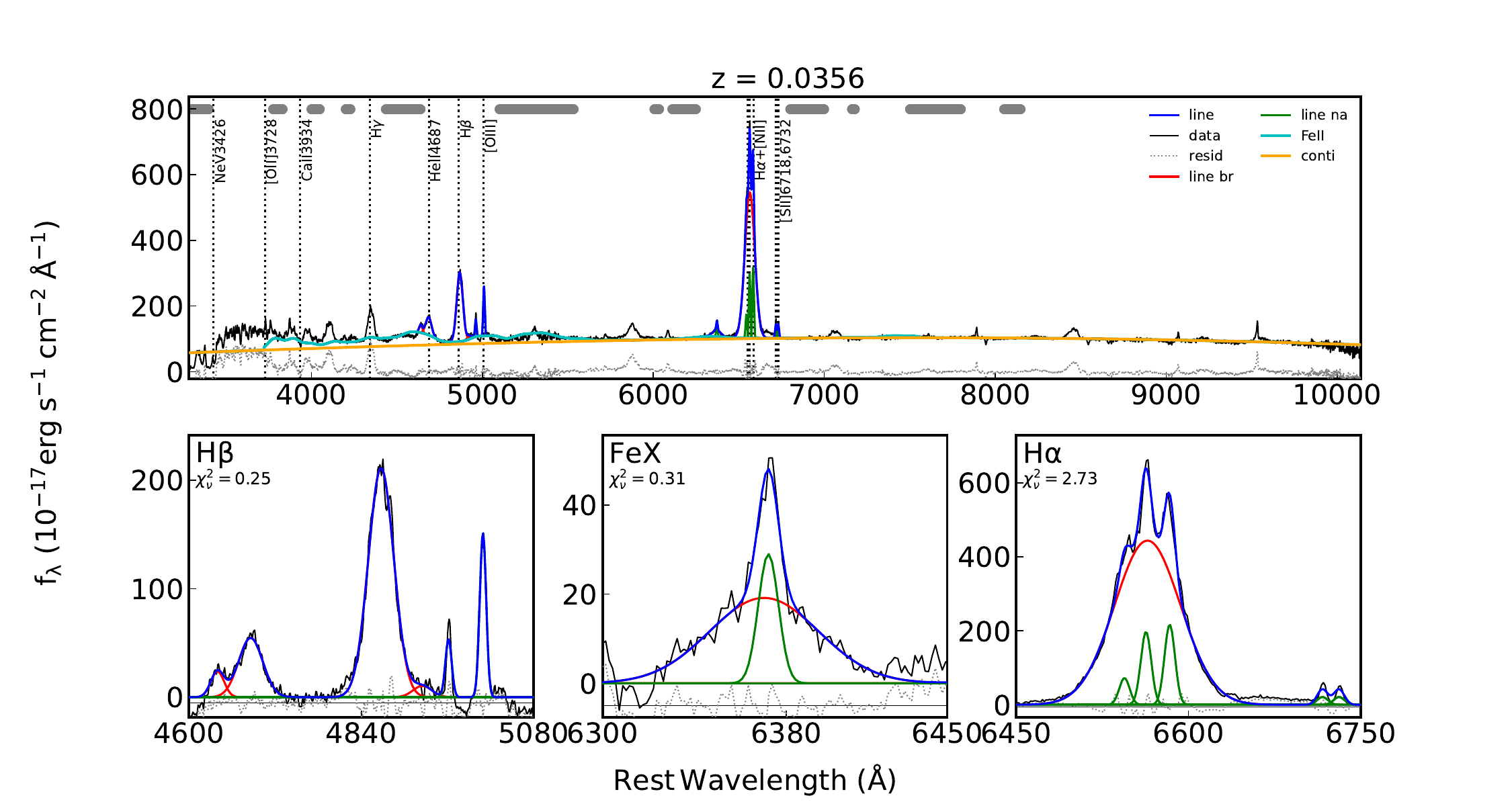}
  \caption{Examples of the extinction-corrected, continuum- and iron-subtracted over-flare-2 spectra fitted using \textsc{PyQSOFit}. The higher-resolution DBSP spectra were convolved to match with LRIS to allow comparison. The emission lines of interest were modeled with a set of broad and/or narrow Gaussian line components. \textit{Left:} The LRIS spectrum taken on 2023-06-15, zooming into the H$\beta$ and H$\alpha$ regions. \textit{Right:} The DBSP spectrum taken on 2024-05-21, where we also fitted the [Fe\,{\sc x}] $\lambda 6375$ high-ionization coronal line as a separate component.}
  \label{fig:a3}
\end{figure*}
\end{document}